\documentclass[letterpaper,twocolumn,10pt]{article}
\usepackage{usenix2019_v3,epsfig,endnotes}

\usepackage{cite}
\usepackage{amsmath,amssymb,amsfonts}
\usepackage[font=footnotesize]{subfig}
\usepackage{graphicx}
\usepackage{textcomp}
\usepackage{xcolor}

\usepackage{custom_macros} 

\begin{document}

\date{}

\title{\Large \bf When the Signal is in the Noise: \\ Exploiting Diffix's Sticky Noise}

\newcommand\CoAuthorMark{\footnotemark[\arabic{footnote}]} 

\author[a]{\rm Andrea Gadotti\footnote{These authors contributed equally.}}
\author[a]{\rm Florimond Houssiau\protect\CoAuthorMark}
\author[a,b]{\rm Luc Rocher\protect\CoAuthorMark}
\author[a]{\rm Benjamin Livshits}
\author[a]{\rm Yves-Alexandre de Montjoye\thanks{Email: \texttt{deMontjoye@imperial.ac.uk}; Corresponding author.\\
We acknowledge the support from Agence Française de Développement. The opinions expressed in this article are the authors' own and do not reflect the view of the Agence Française de Développement.\\
Luc Rocher is the recipient of a doctoral fellowship from the Belgian National Fund for Scientific Research (F.R.S.-FNRS).}}

\affil[a]{\textit{Department of Computing and Data Science Institute, Imperial College London}}
\affil[b]{\textit{ICTEAM, Université catholique de Louvain}}

\maketitle

\begin{abstract}
Anonymized data is highly valuable to both businesses and researchers. A large body of research has however shown the strong limits of the de-identification release-and-forget model, where data is anonymized and shared. This has led to the development of privacy-preserving query-based systems. Based on the idea of ``sticky noise'', Diffix has been recently proposed as a novel query-based mechanism satisfying alone the EU Article~29 Working Party's definition of anonymization. According to its authors, Diffix adds less noise to answers than solutions based on differential privacy while allowing for an unlimited number of queries.

This paper presents a new class of noise-exploitation attacks, exploiting the noise added by the system to infer private information about individuals in the dataset. Our first differential attack uses samples extracted from Diffix in a likelihood ratio test to discriminate between two probability distributions. We show that using this attack against a synthetic best-case dataset allows us to infer private information with 89.4\% accuracy using only 5 attributes. Our second cloning attack uses dummy conditions that conditionally strongly affect the output of the query depending on the value of the private attribute. Using this attack on four real-world datasets, we show that we can infer private attributes of at least 93\% of the users in the dataset with accuracy between 93.3\% and 97.1\%, issuing a median of 304 queries per user. We show how to optimize this attack, targeting 55.4\% of the users and achieving 91.7\% accuracy, using a maximum of only 32 queries per user.

Our attacks demonstrate that adding data-dependent noise, as done by Diffix, is not sufficient to prevent inference of private attributes. We furthermore argue that Diffix alone fails to satisfy Art.~29 WP's definition of anonymization. We conclude by discussing how non-provable privacy-preserving systems {can} be combined with fundamental security principles such as defense-in-depth and auditability to build practically useful anonymization systems without relying on differential privacy.
\end{abstract}
\section{Introduction}

Personal data holds a significant potential for researchers and organizations alike, yet its large-scale collection and use raise serious privacy concerns. 
While scientists have compared the impact of modern large-scale datasets of human behaviors to the invention of the microscope~\cite{Lazer2009-fa}, numerous scandals, such as the recent Cambridge Analytica debacle~\cite{Cadwalladr2018-so} highlight the importance of privacy and data protection for the general public and modern societies.

Historically, the balance between using personal data and preserving people's privacy has relied, both practically and legally, on the concept of data anonymization. Anonymization, also called de-identification, is the process of transforming personal data to mask the identity of participants, e.g.\ by removing identifiers, coarsening data, or adding noise.
The recent European General Data Protection Regulation~(GDPR) defines anonymous information as ``information which does not relate to an identified or identifiable natural person or to personal data rendered anonymous in such a manner that the data subject is not or no longer identifiable''~\cite{Oj_l2016-fi}.
Similar definitions exist in other protection laws around the world, such as the HIPAA privacy rule for medical data in the US and the ePrivacy regulation. These all state that anonymized data does not require consent from participants to be shared widely, as it cannot be traced back and potentially used against them.

While de-identification algorithms are widely used in industry and academia to transform and release anonymous datasets, a large body of research has shown that these practices are not resistant to a wide range of re-identification attacks~\cite{Sweeney1997-cm,Narayanan2008-mj,De_Montjoye2013-sj,De_Montjoye2015-sz,Culnane2017-vc,Ohm2010-zy}. Exposure of the limits of de-identification have led to less than happy conclusions by policy makers: for instance, the [US] President's Council of Advisors on Science and Technology concluded that data anonymization ``is not robust against near-term future re-identification methods. PCAST do not see it as being a useful basis for policy''~\cite{PCAST}.

\point{Query-based systems}
In response to the limits of de-identification, privacy researchers and companies have proposed query-based systems as an alternative. 
Such systems typically offer data analysts a remote interface to ask questions that return data aggregated from several, potentially many, records. Granting access to the data only through queries, without releasing the underlying raw data, mitigates the risk of typical re-identification attacks. Yet a malicious analyst can often submit a series of seemingly innocuous queries whose outputs, when combined, will allow them to infer private information about participants in the dataset.

\point{Differential privacy}
Privacy research has been increasingly focused on providing provable privacy guarantees to defend query-based systems against such attacks. Differential privacy~\cite{diff_priv} is the main privacy guarantee considered by researchers. Intuitively, a randomized algorithm is differentially private if the output does not depend on any single individual's record in the dataset. It has been shown that algorithms that satisfy differential privacy are robust to a very large class of attacks~\cite{Kasiviswanathan2014-am}. 
Yet, efficient differential privacy mechanisms are generally very use case-specific and, even if a large range of differentially private mechanisms have been proposed, there is still no practical widely deployed differential privacy solution for general-purpose data analytics~\cite{uber_flex}. To achieve strong privacy guarantees, practitioners must often sharply limit the data utility by adding large amounts of noise and restrict the total number of requests that the system is allowed to answer~\cite{Tang_Korolova_Bai_Wang_Wang_2017}. Moreover, while differential privacy is a strong guarantee, the risk of data breaches because of implementation issues remains, exposing systems to attacks that differential privacy should in theory rule out \cite{Mironov2012-vh, Haeberlen2011-xr}. Overall, with some rare exceptions, the complexity of correctly implementing differential privacy and choosing the right privacy budget often prevents practitioners from using it.

\point{Alternatives to differential privacy} %
Diffix, a patented commercial solution that acts as an SQL proxy between an analyst and a protected database~\cite{Francis2017-df,Francis_2018-dv}, has recently been proposed by researchers affiliated with Aircloak and the Max Planck Institute for Software Systems as a practical alternative to differential privacy, based on the [EU] Article 29 Working Party (Art. 29 WP)'s definition of anonymous data. It defines a dataset as anonymous if the anonymization mechanism used protects against \emph{singling out} (identify one user), \emph{linkability} (match users across datasets), and \emph{inference} (learn participants' records) attacks~\cite{Art29WP2014-xf,Francis_2018-dv}. Diffix relies on a novel noise addition framework called ``sticky noise'', which aims to give analysts a rich query syntax, minimal noise addition, and an infinite number of queries, all while satisfying the WP29 definition of anonymous.

The authors claim that data produced by Diffix 
($i$) falls outside of the scope of the new European GDPR regulation;  
($ii$) has been determined by the French National Commission on Informatics and Liberty (CNIL) to offer ``GDPR-level anonymization'' for all cases; and 
($iii$) has been certified by TÜViT as fulfilling ``all requirements for data collection and anonymized reporting''~\cite{aircloak2017-ao,aircloak2017-bt}. Diffix is used in production and Aircloak reports working with partners such as Telefonica, DZ Bank and Cisco.

\point{Exploiting Diffix's noise}
In this paper we present a new class of attacks, called noise-exploitation attacks.
The attacks work in three parts: ($i$) canceling out part of the sticky noise using multiple queries, ($ii$) exploiting the noise Diffix adds to one query in order to learn information about the query set associated to this query, and ($iii$) using logical equivalence between queries to obtain independent noise samples for the same query. We develop two noise-exploitation attacks that take advantage of the structure of Diffix's sticky noise to infer private (also called secret) attributes of individuals in the dataset, violating the inference requirement from the Article 29 WP definition of anonymization~\cite{Art29WP2014-xf}. Our first attack, the differential attack, uses samples obtained by the difference between two queries' outputs, to discriminate between two distributions, depending on the value of the private attribute. We show that, under specific conditions, this attack potentially allows an attacker to infer private information of unique users with up to $\theoreticalAccuracyDifferential\%$ accuracy knowing only $\numberOfAttributesDifferential$ pieces of auxiliary information we call attributes.

Our second noise-exploitation attack, the cloning attack, uses dummy conditions that affect the output of queries conditionally to the value of the private attribute. This attack relies on weaker assumptions and automatically validates them with high accuracy, without the need for an oracle. It proceeds in two steps: 
($i$) a \emph{validation} step, searching for subsets of known attributes to use for the attack, that will satisfy the assumptions required for its success, and 
($ii$) an \emph{inference} step that uses the attributes found to predict users’ private attribute's value. 
We perform the attack against four real-world datasets, and show that it can infer the private attributes of between $\minAttacked\%$ and $\maxAttacked\%$ of all records across datasets. We then present an optimized cloning attack that targets 55.4\% of the users and achieves the same accuracy using as little as 32 queries. This proves that introducing limits for the number of allowed queries would not protect against our attacks.

\point{Contributions}
This paper makes the following contributions:
\begin{itemize}

\item By developing and implementing two attacks, we demonstrate that Diffix alone does not currently satisfy the \emph{inference} requirements of the Art.~29 WP. We make the datasets and code for the attacks and experiments available to other researchers.

\item We show, using a collection of four previously published datasets that the assumptions made by our attacks are realistic. We establish that, across all datasets, between 93\% and 100\% of all users are value-unique (i.e.\ all records sharing the same set of attributes have the same private attribute).

\item We make a range of defense-in-depth proposals, which can be used to improve the practical privacy guarantees of both Diffix and other non-differentially private data anonymization tools. While these measures will not result in differential privacy guarantees, they might provide adequate practical solution in many settings.

\item We show, using the Diffix mechanism as our primary example, that anonymization mechanisms that do not rely on differential privacy might not be GDPR-compliant alone, and that naive data-dependent noise can lead to powerful attacks.

\end{itemize}

\point{Paper organization}
The rest of the paper is organized as follows.
Section \ref{sec:description_diffix} describes the Diffix mechanism. 
Section \ref{attack_section} presents two attacks exploiting the noise added by the system to infer private attributes of individuals in the dataset. 
Section \ref{sec:experiments} shows, on real-world datasets, how an attacker can accurately attack the Diffix mechanism.
Section \ref{sec:discussion_section} discusses the assumptions of the attacks and potential solutions for Diffix to thwart noise-exploitation attacks moving forward.
Sections \ref{sec:related_work} and \ref{sec:conclusion} summarize related work and provide our conclusions to build practically useful anonymization systems.
Appendices \ref{LRT-appendix}--\ref{appendix-improvements} provide some details and improvements for the statistical tests used by the attack. Appendix \ref{appendix-number-of-queries} describes how to optimize the cloning attack to reduce the number of queries.
\section{Summary of the Diffix framework}\label{sec:description_diffix}
Here we summarize the Diffix framework as described in \cite{Francis_2018-dv} and introduce notations for our attack. 
Diffix acts as an SQL proxy between an analyst and a \emph{protected database} $D$ where each row is an individual record and each column one attribute. The analyst can send SQL queries to Diffix, which will process the queries and then output a noisy answer.

We denote with $\A_D$ the set of attributes in the database $D$. For instance, $\A_D$ could contain 4 attributes $\A_D = \{\textit{gender}, \textit{age}, \textit{zip}, \textit{HIV}\}$ with $\textit{HIV}$ a binary attribute (0 or 1). A record $x$ is a row of $D$ with values for the attributes in $\A_D$. For example, with $\A_D$ as above, we could have $x = (M, 27, 55416, 1)$. We assume, for simplicity, that there is one and only one record for every user in $D$.

While Diffix can process a large part of the SQL syntax, we here focus on simple count queries:
\begin{align*}
& \textsf{SELECT count}(*)\\
& \textsf{FROM } table\\
& \textsf{WHERE } condition_1 \textsf{ AND } condition_2 \textsf{ [AND \ldots]}
\end{align*}

\noindent where every condition is an expression of the form ``$attribute~\square~value$'' with $\square$ being $=$, $\neq$, $\leq$, $<$, $\geq$, or $>$.

\noindent For simplicity, we use a shorter notation for queries using ``$\wedge$'' for the logical \textsf{AND}:
\[
Q \equiv \coun(condition_1 \wedge condition_2 \wedge \ldots).
\]
The following query would, for example, count the number of individuals in the database who are male, 37 years old, and live in the area with ZIP code 48828:
\[
Q \equiv \coun(gender=M \wedge age=37 \wedge zip=48828)
\]

\begin{figure}[bt]
	\centering
	\includegraphics[width=.9\columnwidth]{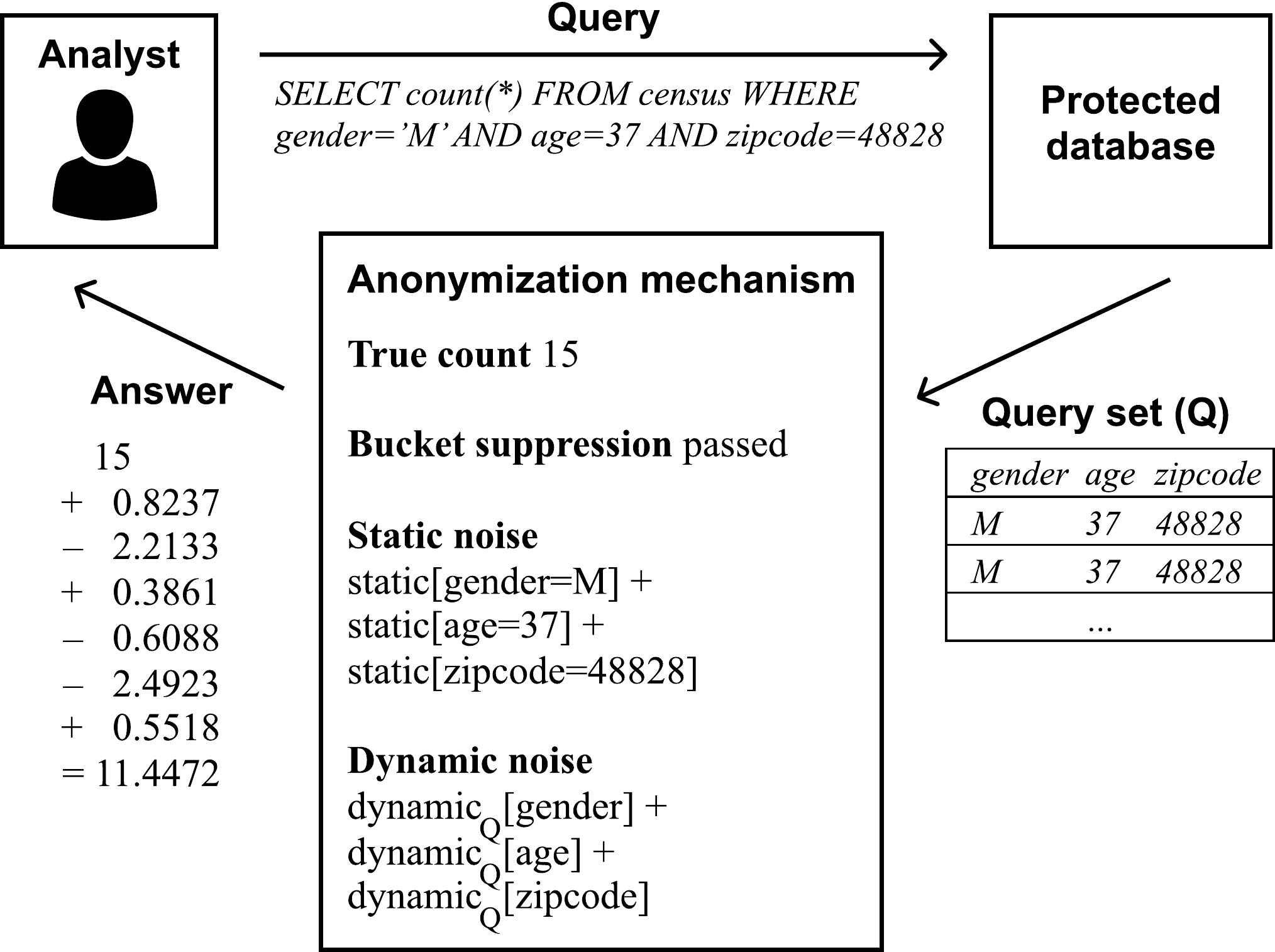}
    \caption{Diffix privacy-preserving architecture}
  	\label{img:diffix-system}
\end{figure}

\point{Diffix's privacy-preserving mechanism}
Diffix protects privacy through \emph{sticky noise} addition (static and dynamic noise) and bucket suppression (see Fig.~\ref{img:diffix-system}). Let $Q \equiv \coun(C_1 \wedge \ldots \wedge C_h)$, and denote by $Q(D)$ the true result of $Q$ evaluated on $D$ (without noise). Diffix's output for $Q$ on $D$ (without bucket suppression, see below) is:
\begin{align}
\widetilde{Q}(D) = Q(D) + \sum_{i=1}^h \static[C_i] + \sum_{i=1}^h \dynamic_Q[C_i] \label{eqn_output}
\end{align}
with $\static[C_i]$ the \emph{static noise} for condition $C_i$, $\dynamic_Q[C_i]$ the \emph{dynamic noise} for condition $C_i$ in $Q$.
\point{Static noise}
Let $C$ be a condition, for example $age=34$. The \emph{static noise} $\static[C]$ associated to $C$ is a random number drawn from a normal distribution $\N(0,1)$. The value is generated by a pseudo-random number generator (PRNG), whose seed is a salted hash of the string literal $C$:
\[
static\_seed_C = \XOR(\hash(C), salt)
\]
This ensures that the static noise associated with $C$ is always the same independently of the query where $C$ appears. The noise is ``sticky'' thereby preventing an attacker to send the same query multiple times, average out the results, and obtain a precise estimate of the private value (\textit{averaging attack})~\cite{Francis_2018-dv}.

\point{Dynamic noise}
In the Diffix framework, every record in $D$ is associated with a user ID, a unique string for that user. These pseudonyms are used to compute the dynamic noise. Let $Q \equiv \coun(C_1 \wedge \ldots \wedge C_h)$ be a query and $C$ any condition $C_i$. The dynamic noise depends not only on $C$, but also on the \emph{query set} of $Q$ in the dataset $D$, i.e.\ the set of users which satisfies \emph{all} conditions $C_1,\ldots,C_h$. More precisely, if the query set for $Q$ on $D$ is $S = \{uid_1, uid_2,\ldots,uid_m\}$, the dynamic noise for $C$ ($\dynamic_Q[C]$) is generated from a normal distribution $\N(0,1)$ by the PRNG seeded with:
\begin{align*}
dynamic\_seed =\XOR(&static\_seed_C, \\
&\hash(uid_1), \;\ldots,\; \hash(uid_m))
\end{align*}
Note that we don't include $D$ in the notation $\dynamic_Q[C]$, as the dataset is usually fixed and clear from the context.

The output $\widetilde{Q}(D)$ is therefore the realization of a random variable distributed as a normal distribution $\N(\mu,\sigma^2)$, with mean $\mu = Q(D)$ and variance $\sigma^2 = 2h$.

\point{Example}
Consider again the query
\[
Q \equiv \coun({gender=M} \wedge {age=37} \wedge {zip=48828}).
\]
Diffix's output for $Q$ on the database $D$ is
\begin{align*}
\widetilde{Q}(D) = \quad & Q(D) \\
+ & \static[{gender=M}] + \dynamic_Q[{gender=M}] \\
+ & \static[{age=37}] + \dynamic_Q[{age=37}] \\
+ & \static[{zip=48828}] + \dynamic_Q[{zip=48828}]
\end{align*}
where $\widetilde{Q}(D)$ is a random value drawn from a normal distribution $\N(Q(D),6)$.

Static and dynamic noise layers are both needed to prevent \emph{intersection attacks}~\cite{Francis_2018-dv,Francis2017-df}, a class of attacks that combine multiple queries to infer private attributes of records.

\point{Bucket suppression}
In addition to static and dynamic noise, Diffix implements another security measure called \emph{bucket suppression}, similar to the classic query set size restriction. If the size of the query set is smaller than a certain threshold, the bucket suppression rejects the query. Previous research has shown that a fixed threshold constitutes a risk for privacy~\cite{stallings}. Diffix addresses this issue by using a \emph{noisy} (and sticky to the query set) \emph{threshold}. Specifically, suppose Diffix processes a query $Q \equiv \coun(C_1 \wedge \ldots \wedge C_h)$. If $Q(D) \leq 2$, then the query gets suppressed. If $Q(D) > 1$, then Diffix computes a noisy threshold $T \sim \N(4, 1/2)$, using the seed:
\[
\mathit{threshold\_seed} = \XOR(salt, \hash(uid_1), \ldots, \hash(uid_m))
\]

If $Q(D) < T$, the query is suppressed; otherwise, the noisy output $\widetilde{Q}(D)$ is computed and sent to the analyst. In the original Diffix mechanism~\cite{Francis2017-df}, the queries are said to be ``silently suppressed'' when censored by bucket suppression. This could mean that (1) a value of 0 is returned as result, (2) a random value is returned or (3) Diffix displays an error message.
In this paper, we assume that a bucket-suppressed query will return a value of zero. This gives less information to a potential attacker than an error message, as a value of zero can be the result of either noise addition or bucket suppression. We consider that returning a random result affects utility too significantly to be applied in practice.
\section{Noise-exploitation attacks}
\label{attack_section}
Our noise-exploitation attacks, which we call \textit{differential} and \textit{cloning}, are both based on three observations. First, since the noise is sticky, it is possible to \emph{cancel out} part of it using multiple queries. Second, since the noise depends on the query set, the noise itself leaks information about the query set. Third, exploiting logical equivalence between some queries, it is possible to circumvent the ``stickiness'' of the noise by repeating (almost) the same query and consequently obtain independent noise samples. Our differential attack uses samples in a likelihood ratio test to discriminate between two probability distributions depending on the value of the private (also called secret) attribute. Our cloning attack relies on dummy conditions that conditionally strongly affect the output of the query depending on the value of the secret attribute.

\subsection{Differential noise-exploitation attack}

We first define further notations: with $A \subseteq \A_D$ a set of attributes, $x^{(A)}$ is the \emph{restriction} of the record $x$ to $A$, i.e.\ the vector one obtains after removing from $x$ every entry for attributes that are not in $A$. For example, if $\A_D = \{\textit{gender}, \textit{age}, \textit{zip}, \textit{HIV}\}$, $x = (M, 27, 55416, 1)$ and $A = \{\textit{gender}, \textit{age}, \textit{HIV}\}$, then $x^{(A)} = (M, 27, 1)$. If $A$ contains a single attribute $a$, we simply write $x^{(a)}$. So, for example, $x^{(gender)} = M$.

For this attack, we make the following assumptions:

\begin{itemize}
\item[H1] The attacker wants to find out some information about Bob, the victim. The attacker knows that Bob's record is in the dataset. We denote Bob's record by $x$. The attacker has access to the protected dataset only through Diffix.
\item[H2] The attacker knows all of Bob's attributes for some set of attributes $A$. Our \emph{background knowledge} (also called auxiliary information in the literature) is the restricted record $x^{(A)}$. This is a standard assumption in the literature on re-identification attacks \cite{adamSecuritycontrolMethodsStatistical1989, samaratiProtectingPrivacyWhen1998a}.
\item[H3] The attacker wants to infer a secret attribute $s$ about Bob, the victim. That is, she wants to infer the value of $x^{(s)}$, where $s \not\in A$. For simplicity of notation, $s$ is a binary attribute, i.e.\ $x^{(s)} \in \{0,1\}$. This means that the attack can be seen as a classifier, with the output of the attack being negative if the algorithm returns $x^{(s)} = 0$ and positive if it returns $x^{(s)} = 1$.\\
While we here focus on the binary case, our results fairly easily extend to non-binary cases.
\item[H4] There exists an oracle $\Unique$ that takes as input any restricted record $z^{(R)}$ and outputs whether $z^{(R)}$ is \emph{unique}. $\Unique(z^{(R)}) = \text{True}$ if and only if there is no other record $y$ in $D$ such that $z^{(R)} = y^{(R)}$.
\end{itemize}

For the attack to succeed, we first need to ensure that the background knowledge $x^{(A)}$ uniquely identifies Bob. This is given to us by the oracle $\Unique(x^{(A)})$. In this attack, we only attempt to infer secret attributes of records that are unique in the dataset. The cloning attack, presented later, extends this by requiring a weaker notion of uniqueness, which we call value-uniqueness. 

While the existence of such an oracle is a strong assumption, it is weaker than Diffix's claims to protect against an ``analyst [that] has full access to the inference dataset''~\cite{Francis_2018-dv}, where the \emph{inference dataset} is the original dataset with only the secret attribute $x^{(s)}$ removed. If the attacker has access to every record, she can easily verify that no other record shares the same restricted record $x^{(A)}$.

Firstly, the attack needs to bypass bucket suppression. For example, an attacker could ask how many records have both the background knowledge $x^{(A)}$ and the private attribute $s = 0$:
\begin{equation}
Q \equiv \coun(a_1 = x_1 \wedge \ldots \wedge a_k = x_k \wedge s = 0). \label{eqn_Q}
\end{equation}
with $A = \{a_1, \ldots, a_k\}$ and $x^{(A)} = (x_1, \ldots, x_k)$.

While an accurate answer to $Q$ would immediately disclose the value of $x^{(s)}$, since $Q(D)$ can be either $0$ (if $x^{(s)} = 1$) or $1$ (if $x^{(s)} = 0$), this query will always be blocked by bucket suppression since the query set is either empty or $\{\text{Bob}\}$.

\emph{Intersection attacks} have been proposed in the literature to circumvent similar kinds of restrictions \cite{Beck_1980}. Picking an attribute, e.g.\ $a_1$, we can define the following queries:
\begin{align}
Q_1 & \equiv \coun(a_2 = x_2 \wedge \ldots \wedge a_k = x_k \wedge s = 0) \label{eqn_Q1} \\
Q'_1 & \!\begin{multlined}[t][0.8\displaywidth]
\equiv \coun(a_1 \neq x_1 \wedge a_2 = x_2 \wedge \ldots \\
\ldots \wedge a_k = x_k \wedge s = 0)
\end{multlined} \label{eqn_Q2}
\end{align}
As the record $x$ is unique, by assumption, it is the only record that can differ between $Q_1$ and $Q'_1$. This allows us to directly compute $Q(D)$:
\begin{align}
Q(D) = Q_1(D) - Q'_1(D) = \begin{cases}
  0 \quad & \text{ if } x^{(s)} = 1 \\
  1 \quad & \text{ if } x^{(s)} = 0
\end{cases} \label{eqn_Q12}
\end{align}

To prevent this vulnerability\footnote{For the intersection attack to be successful, both $Q_1$ and $Q'_1$ need to be large enough to not trigger the bucket suppression. We discuss this assumption later on.}, Diffix adds static and dynamic noises:
\begin{align*}
\widetilde{Q}_1(D) =\quad &Q_1(D) \\
+ &\sum_{i = 2}^k \static[a_i = x_i] + \static[s = 0] \\
+ &\sum_{i = 2}^k \dynamic_{Q_1}[a_i = x_i] + \dynamic_{Q_1}[s = 0]
\end{align*}
and
\begin{align*}
\widetilde{Q}'_1(D) =\quad &Q'_1(D) \\
+ &\static[a_1 \neq x_1] + \dynamic_{Q'_1}[a_1 \neq x_1]\\
+ &\sum_{i = 2}^k \static[a_i = x_i] + \static[s = 0] \\
+ &\sum_{i = 2}^k \dynamic_{Q'_1}[a_i = x_i] + \dynamic_{Q'_1}[s = 0].
\end{align*}

The first part of our attack relies on noticing that $k-1$ static noise layers cancel out. Let
\[
q_1 = \widetilde{Q}_1(D) - \widetilde{Q}'_1(D).
\]
Then:
\begin{align*}
q_1 =\quad &Q_1(D) - Q'_1(D) \\
- &\static[a_1 \neq x_1] - \dynamic_{Q'_1}[a_1 \neq x_1] \tag{\textit{fixed}} \\
+ &\sum_{i = 2}^k \dynamic_{Q_1}[a_i = x_i] + \dynamic_{Q_1}[s = 0] \tag{$\textit{dynamic}_{Q_1}$} \\
- &\sum_{i = 2}^k \dynamic_{Q'_1}[a_i = x_i] - \dynamic_{Q'_1}[s = 0] \tag{$\textit{dynamic}_{Q'_1}$}
\end{align*}
leaving us with $2k+2$ noise layers: $\textit{fixed} \sim \N(0,2)$, $\textit{dynamic}_{Q_1} \sim \N(0,k)$, and $\textit{dynamic}_{Q'_1} \sim \N(0,k)$.

The second part of the attack relies on the fact that both $\dynamic_{Q_1}[a_i = x_i]$ and $\dynamic_{Q'_1}[a_i = x_i]$ (resp. $\dynamic_{Q_1}[s = 0]$ and $\dynamic_{Q'_1}[s = 0]$) relate to the same condition ``$a_i = x_i$'' (resp. ``$s = 0$''). This means that the noise values for $Q_1$ and $Q'_1$ will cancel out if $Q_1$ and $Q'_1$ have the same query set. Therefore either $Q_1(D)-Q'_1(D) = 0$, and the $2k$ dynamic noise layers cancel out, or $Q_1(D)-Q'_1(D) = 1$, and no dynamic noise layer is canceled out:
\begin{align}
q_1 \sim \begin{cases}
  \N(0,2) \quad & \text{ if } x^{(s)} = 1 \\
  \N(1,2k+2) \quad & \text{ if } x^{(s)} = 0
\end{cases}
\label{eq:distribution-cases}
\end{align}

Using this result, an attacker can run a likelihood ratio test (see Appendix \ref{LRT-appendix}) to estimate whether $q_1$ is distributed as $\N(0,2)$ or $\N(1,2k+2)$ and predict the value of $x^{(s)}$. The larger $k$ is, the easier it becomes to discriminate between the two distributions. This alone already allows the attacker to infer Bob's secret, $x^{(s)}$, with better than random accuracy.

The third part of the attack allows us to strongly improve the accuracy of our inference. While the stickiness of the noise prevents us from running the same query again to collect more sample, we circumvent it by using different pairs of queries for which equation (\ref{eq:distribution-cases}) is still true. Specifically, instead of removing (resp.\ negating) the condition $a_1 = x_1$, we remove (resp.\ negate) other conditions $a_j = x_j$ for $j \leq k$, obtaining queries $Q_j$ (resp.\ $Q'_j$) for $j \leq k$. In our notation:
\begin{align}
Q_j & \equiv \coun\left( \bigwedge_{\substack{i=1 \\ i \neq j}}^{k} a_i = x_i \wedge s = 0\right) \label{eq:Qj} \\
Q'_j & \equiv \coun\left( \bigwedge_{\substack{i=1 \\ i \neq j}}^{k} a_i = x_i \wedge a_j \neq x_j \wedge s = 0\right) \label{eq:Q'j}
\end{align}
Running all queries $\{(Q_j,Q'_j)\}_{j \leq k}$, the attacker collects a vector of realizations $\{q_j\}_{j \leq k}$ where $q_j = \widetilde{Q}_j(D) - \widetilde{Q}'_j(D)$. All $q_j$ values are computed from different queries, which generate different noises. Hence, the noises all have different values (with probability 1). Nevertheless, the equation~(\ref{eq:distribution-cases}) is still true for each $q_j$, so we can combine them as $k$ different samples from the same distribution, and estimate the value of $Q(D)$.

Finally, replacing the ``$s = 0$'' condition with ``$s = 1$'' in every pair $(Q_j,Q'_j)$ defines $k$ new pairs of queries $\{(R_j,R'_j)\}_{j \leq k}$. This allows us to obtain $k$ more samples $\{r_j\}_{j \leq k}$ (with different noises and inverted results in equation~(\ref{eq:distribution-cases})). This gives us a total of $2k$ samples before bucket suppression (see Appendix \ref{LRT-appendix}), generated by issuing $4k$ queries.

On a technical note, observe that in principle we cannot be certain that the $q_j$'s (resp.\ $r_j$'s) are all independent samples, because two different queries $Q_j$ and $Q_l$ (resp.\ $R_j$ and $R_l$) with $j \neq l$ might have the same query set. In that case, the dynamic noise layers associated to the same conditions would have the same values, and hence the total dynamic noises of the two queries would be heavily correlated. While this affects the accuracy of the likelihood ratio test, the impact is negligible in practice. Hereafter, we always assume that the samples are independent.

\point{Example} We present an example of how the differential attack would work in practice. We consider a dataset $D$ containing the attributes $age$, $zip$, and $\textit{HIV}$, and assume that the attacker knows Bob's restricted record $x \setminus x^{(\textit{HIV})}$ for which $x^{(age)} = 37$ and $x^{(zip)} = 48828$. As before, we assume that Bob is the only user with both $age=37$ and $zip=48828$. Consider the query:
\begin{equation*}
Q \equiv \coun({age=37} \wedge {zip=48828} \wedge {\textit{HIV}=0}).
\end{equation*}
which has true value $0$ if Bob has \textit{HIV}, and $1$ if Bob does not have HIV. To circumvent the bucket suppression, the attacker defines two queries for the intersection attack. The first pair of queries removes (resp.\ negates) the $age$ condition:
\begin{align*}
Q_1 & \equiv \coun({zip=48828} \wedge {\textit{HIV}=0}) \\
Q'_1 & \equiv \coun({age}\neq37 \wedge {zip=48828} \wedge {\textit{HIV}=0}).
\end{align*}

Based on the previous findings, we know that $q_1 = \widetilde{Q}_1(D) - \widetilde{Q}'_1(D)$ satisfies:
\begin{align}
q_1 \sim
\begin{cases}
\N(0,2) & \quad \text{if Bob has HIV} \\
\N(1,4) & \quad \text{if Bob does not have HIV}.
\end{cases}
\end{align}
The realization $q_1$ is a first sample for our likelihood ratio test. Repeating the same procedure with the $zip$ attribute gives us:
\begin{align*}
Q_2 & \equiv \coun({age} = 37 \wedge {\textit{HIV}=0}) \\
Q'_2 & \equiv \coun({age} = 37 \wedge {zip}\neq48828 \wedge {\textit{HIV}=0}),
\end{align*}
and $q_2 = \widetilde{Q}_2(D) - \widetilde{Q}'_2(D)$ is again distributed as $\N(0,2)$ if Bob has HIV, and as $\N(1,4)$ otherwise. This is the second sample. Repeating the procedure defining queries $R_1$ and $R'_1$, as well as $R_2$ and $R'_2$, and replacing $\textit{HIV}=0$ with $\textit{HIV}=1$ (and inverting the results), the attacker obtains two more samples.

\point{Full differential attack}
For larger values of $k$, the queries used by the differential attack contain many conditions, and hence potentially select a low number of records. Depending on the dataset, this might result in a large fraction of queries being bucket suppressed, leaving the attacker with few or no samples for the likelihood ratio test. To counteract this effect, we integrate the attack with a subset exploration step to obtain a \emph{full differential attack}. Assume that the attacker knows a set $A^*$ of the correct attributes for the victim with $|A^*|=k^*$, i.e.\ the background knowledge is $x^{(A^*)}$. The full attack proceeds as follows. The algorithm selects random subsets of $A^*$ until it finds a subset $A \subseteq A^*$ such that $\Unique(x^{(A)})$ returns True. It then performs the differential attack using $x^{(A)}$ as background knowledge. For the likelihood ratio test, the attack considers only query pairs $(\widetilde{Q}_j(D), \widetilde{Q}'_j(D))$ or $(\widetilde{R}_j(D), \widetilde{R}'_j(D))$ which have outputs larger than zero in both entries. If no such pair exists, the algorithm samples a new subset $A$ and iterates the procedure. If no feasible subset is found, the algorithm outputs NonAttackable. The subsets of $A^*$ are sampled by decreasing size, as bucket suppression is less likely for lower values of $k$ (but on the other hand the likelihood ratio test is less accurate).

The procedure \ref{proc:full-differential-attack} presents in detail the algorithm outlined above.

\begin{procedure}
    \caption{DifferentialAttack($A, x^{(A)}, s$)}\label{proc:differential-attack}
	\KwIn{known attributes (names $A$ and values $x^{(A)}$), secret $s$}
	\KwOut{True if $x^{(s)} = 1$, False if $x^{(s)} = 0$, NoSamples if $x^{(A)}$ does not yield any positive sample}
    $k \gets |A|$, \ $\Q \gets \emptyset$, \ $\R \gets \emptyset$ \;
    \For{$j \gets 1$ \KwTo $k$}{
        $I \gets \{ i \in  [1, k] \;|\; i \neq j \}$ \;
        $\widetilde{Q} \gets \coun\left( \bigwedge_{i \in I } a_i = x_i \wedge s = 0\right)$ \;
        $\widetilde{Q}' \gets \coun\left( \bigwedge_{i \in I} a_i = x_i \wedge a_j \neq x_j \wedge s = 0\right)$ \;
     	  \If{$\widetilde{Q}>0$ \normalfont{and} $\widetilde{Q}'>0$}{
     	      $q_j \gets \widetilde{Q} - \widetilde{Q}'$ \;
     	      $\Q \gets \Q \cup \{q_j\}$
     	  }
    }
    \For{$j \gets 1$ \KwTo $k$}{
        $I \gets \{ i \in  [1, k] \;|\; i \neq j \}$ \;
        $\widetilde{R} \gets \coun\left( \bigwedge_{i \in I } a_i = x_i \wedge s = 1\right)$ \;
        $\widetilde{R}' \gets \coun\left( \bigwedge_{i \in I } a_i = x_i \wedge a_j \neq x_j \wedge s = 1\right)$ \;
     	  \If{$\widetilde{R}>0$ \normalfont{and} $\widetilde{R}'>0$}{
     	      $r_j \gets \widetilde{R} - \widetilde{R}'$ \;
     	      $\R \gets \R \cup \{r_j\}$
     	  }
    }
    \If{$\Q = \emptyset$ \normalfont{and} $\R = \emptyset$}{\Return \normalfont{NoSamples}}
    $f \gets$ PDF of $\N(0,2)$,\quad $g \gets$ PDF of $\N(1,2k+2)$ \;    
    $L \gets \prod_{q \in \Q} \frac{f(q)}{g(q)} \prod_{r \in \R} \frac{g(r)}{f(r)}$ \;
    \Return $L \geq 1$
\end{procedure}

\begin{procedure}
    \caption{FullDifferentialAttack($A^*, x^{(A^*)}, s$)}\label{proc:full-differential-attack}
	\KwIn{known attributes (names $A^*$ and values $x^{(A^*)}$), secret $s$}
	\KwOut{True if $x^{(s)} = 1$, False if $x^{(s)} = 0$, NonAttackable if $x^{(A^*)}$ is not attackable}
  \For{$k \gets |A^*|$ \KwTo $1$}{
   	\For{iter $\gets 1$ \KwTo $100$}{
 		  $A \gets \text{RandomSubsetOfSize}(A^*, k)$ \;
  	  \If{{\upshape$\Unique(x^{(A)})$ and \ref{proc:differential-attack}$(A, x^{(A)}, s) \neq$ NoSamples}}{
  	    \Return \ref{proc:differential-attack}$(A, x^{(A)}, s)$
  	  }
	  }
	}
  \Return NonAttackable	  
\end{procedure}

\subsection{Cloning noise-exploitation attack}

In this section, we present an extension of the differential noise-exploitation attack, which we call \emph{cloning attack}. This attack adds dummy conditions, that don't affect the query set, to queries, in such a way that several queries with different dummy conditions will have either identical or very different results conditional to the secret attribute's value.

We first introduce some new notations and definitions. Denoting by $x$ the victim's entire record, the attacker's background information is now $x^{(A)} = (x^{(A')},x^{(u)})$ with $A = A' \cup \{u\}$ and $|A|=k$. We use the shorthand $(A',u)$ for $A' \cup \{u\}$. We also define a restricted record $z^{(A)}$ to be \emph{value-unique} for the attribute $s$ in $D$ if $y^{(A)} = z^{(A)}$ implies $y^{(s)} = z^{(s)}$. That is, every record that shares the same attributes of the restricted record $z^{(A)}$ holds the same value for the secret attribute $s$. To simplify the notation, we write $A' = x^{(A')}$ to indicate the condition that all attributes in $A'$ must match the ones in $x^{(A')}$, i.e.\ $\wedge_{a \in A'} a = x^{(a)}$

The attack addresses several limitations of the differential attack, making it much stronger in practice.

First, the cloning attack does not require an oracle to confirm that the background information uniquely identifies a user. Instead, it replaces the oracle with a heuristic to automatically validate the assumptions.

Second, the differential attack has to ignore any pair $(\widetilde{Q}_j(D), \widetilde{Q}'_j(D))$ where at least one of the entries is not positive, as it cannot tell whether the null output comes from noise addition or from bucket suppression. This significantly reduces the total number of samples used. Our cloning attack instead uses ``dummy conditions'' that do not impact the user set. As queries now only differ in the dummy conditions, the corresponding query sets will always be identical. This allows us to rule out bucket suppression in case at least one output is greater than zero.

Third, while the differential attack requires records to be unique, the cloning attack only requires records to be value-unique. This is a weaker condition and makes the cloning attacks effective on a larger set of users.

Fourth, the cloning attack only requires that the set of users who share all attributes in $x^{(A')}$ is ``large enough'' to not be bucket suppressed. This is a weaker assumption than for the differential attack, where this needed to hold for a large number of subsets of $A$ with one attribute removed. Furthermore, the cloning attack validates this automatically (and thus prevents bucket suppression) with high confidence.

While much stronger, the attack relies on the attacker being able to produce a set of distinct \emph{dummy conditions} $\Delta = \{\Delta_j\}_{1 \leq j \leq |\Delta|}$, where each $\Delta_j$ is an SQL statement such that the set of users matching $A' = x^{(A')}$ is the same as the set of users matching $A' = x^{(A')} \land \Delta_j$. In section~\ref{sec:discussion_section}, we discuss how dummy conditions are easy to obtain, slow to detect, and how automatically filtering them might introduce new vulnerabilities.

\point{Description of the attack}
For each dummy condition $\Delta_j$, we define the two queries:
\begin{align}
Q_j & \equiv \coun\left(A' = x^{(A')} \wedge \Delta_j \wedge s = 0\right) \label{eqn_Q1_atk2} \\
Q'_j & \equiv \coun\left(A' = x^{(A')} \wedge \Delta_j \wedge u \neq x^{(u)} \wedge s = 0\right) \label{eqn_Q2_atk2}
\end{align}

With $q_j = \widetilde{Q}_j(D) - \widetilde{Q}'_j(D)$, we have:
\begin{align*}
q_j = \quad & Q_j(D) - Q'_j(D) \\
- & \static[u \neq x^{(u)}] ~- \dynamic_{Q_j}[u \neq x^{(u)}] \\
+ & \sum_{i \in A'} \dynamic_{Q_j}[a^{(i)} = x^{(i)}] + \dynamic_{Q_j}[s=0] \\
- & \sum_{i \in A'} \dynamic_{Q'_j}[a^{(i)} = x^{(i)}] - \dynamic_{Q'_j}[s=0] \\
+ & \left(\dynamic_{Q_j}[\Delta_j] - \dynamic_{Q'_j}[\Delta_j]\right)
\end{align*}

By the same argument we presented for the differential attack, if $x^{(s)} = 1$ then $Q_j(D) = Q'_j(D)$ and most dynamic and static noises cancel out, giving:
\begin{equation}
q_j = - \static[u \neq x^{(u)}] - \dynamic_{Q_j}[u \neq x^{(u)}]
\end{equation}

As this value does not depend on the dummy condition used, we have that $q_1 = q_2 = \dots = q_{|\Delta|}$.

On the contrary, if $x^{(s)} = 0$, then the noise layers do not cancel out with probability 1. As the noise values given by $\dynamic_{Q_j}[\Delta_j]$ and $\dynamic_{Q'_j}[\Delta_j]$ depend on $\Delta_j$, the probability that all (or any) $q_j$ are equal is zero.

We can therefore complete the attack by inferring that $x^{(s)} = 1$ if $q_1 = \dots = q_{|\Delta|}$, and $x^{(s)} = 0$ otherwise. Under the current assumptions, the attack always infers the correct value with 100\% confidence (up to pseudo-random collisions in Diffix's noise addition mechanism).

\point{Robustness against rounding}
In the previous section, we follow the Diffix papers~\cite{Francis_2018-dv, Francis2017-df} and assume that $\widetilde{Q}(D)$ is returned directly without any rounding, admitting also negative values.
We now consider the case where results are rounded to the nearest nonnegative integer, and propose a simple modification of our attack that accounts for this. 

When the results of the queries $Q_j$ and $Q'_j$ are rounded, the corresponding $q_j$ might not be identical if $x^{(s)} = 0$. However, the $q_j$s will vary less if $x^{(s)}=1$ than if $x^{(s)}=0$.
Hence, instead of checking if $q_1 = \ldots = q_{|\Delta|}$, we check if the $q_j$ values are ``similar'' to one another. While for high values of $k$ this is easy to detect, the total variance of the noise for low values of $k$ is small, making it harder to distinguish between the two hypotheses (i.e.\ whether the $q_j$ values are ``similar'' or not). To overcome this issue, we ``amplify'' the noise for each query: instead of adding a single dummy condition $\Delta_j$ to the queries for $Q_j$ and $Q'_j$, we add the conjunction $\wedge_{l \neq j} \Delta_l$:
\begin{align}
Q_j & \equiv \coun\left(A' = x^{(A')} \wedge \bigwedge_{l \neq j} \Delta_l \wedge s = 0\right) \label{eqn_Q1_atk2_new} \\
Q'_j & \equiv \coun\left(A' = x^{(A')} \wedge \bigwedge_{l \neq j} \Delta_l \wedge u \neq x^{(u)} \wedge s = 0\right) \label{eqn_Q2_atk2_new}
\end{align}

This increases the total variance of the noise in $q_j$ in the $x^{(s)}=0$ case, making it easy to distinguish between the two hypotheses: all the $q_j$ values will be very similar if $x^{(s)}=1$ and fluctuate heavily if $x^{(s)}=0$. Measuring the sample variance $S^2$ of $\{q_j\}_{1 \leq j \leq |\Delta|}$, we infer that $x^{(s)} = 1$ if $S^2 \leq \sigma^*$, and $x^{(s)} = 0$ otherwise with a cutoff threshold $\sigma^*$ chosen by the attacker (see Appendix \ref{threshold-appendix} for an empirical analysis of $\sigma^*$). The cloning attack is described in detail in the procedure \ref{proc:dummy-cloning-attack}.

\begin{procedure}
    \caption{CloningAttack($A', u, x^{(A',u)}, \Delta, s, v$)}\label{proc:dummy-cloning-attack}
	\KwIn{known attributes (names $A', u$ and values $x^{(A',u)}$), dummy conditions $\Delta$, secret $s$ and target value $v$}
	\KwOut{True if $x^{(s)} = v$, False if $x^{(s)} \neq v$}
    \For{$j \gets 1$ \KwTo $|\Delta|$}{
    $\varphi \gets A' = x^{(A')} \wedge \bigwedge_{l \neq j} \Delta_l$ \;
    $\widetilde{Q} \gets \coun\left(\varphi \wedge s \neq v\right)$ \;
    $\widetilde{Q}' \gets \coun\left(\varphi \wedge u \neq x^{(u)} \wedge s \neq v\right)$ \;
    $q_j \gets \widetilde{Q} - \widetilde{Q}'$
    }
    $\overline{r} \gets \frac{1}{|\Delta|} \sum_{j=1}^{|\Delta|} q_j$,\quad $S^2 \gets \frac{1}{|\Delta|-1} \sum_{j=1}^{|\Delta|} (q_j - \overline{r})^2$ \;
    \Return $S^2 \leq \sigma^*$
\end{procedure}

\point{Automated validation of the assumption}
The cloning attack relies on two assumptions on the attacker's background knowledge $x^{(A',u)}$:
\begin{enumerate}
\item The queries $\{Q_j\}_{1 \leq j \leq |\Delta|}$ and $\{Q'_j\}_{1 \leq j \leq |\Delta|}$ in equations (\ref{eqn_Q1_atk2_new}) and (\ref{eqn_Q2_atk2_new}) are not bucket suppressed.
\item The user is value-unique in the dataset according to $(A',u)$ for the secret attribute $s$.
\end{enumerate}

We here propose procedures for an attacker to determine whether $(A',u)$ satisfies the two assumptions with high probability.

Validating the first assumption can be done easily by submitting queries $\{Q_j\}_{1 \leq j \leq |\Delta|}$ and $\{Q'_j\}_{1 \leq j \leq |\Delta|}$ to Diffix. Recall that the threshold for bucket suppression for a query depends only on the corresponding query set. All the queries in $\{Q_j\}_{1 \leq j \leq |\Delta|}$ have the same query set, and the same applies for $\{Q'_j\}_{1 \leq j \leq |\Delta|}$. Hence, if \emph{any} query $Q_j$ is bucket suppressed (i.e.\ has output zero), then \emph{all} queries in $\{Q_j\}_{1 \leq j \leq |\Delta|}$ must have output zero, and similarly for $\{Q'_j\}_{1 \leq j \leq |\Delta|}$. Thus, if \emph{any} query $Q_j$ and \emph{any} query $Q'_j$ have output higher than zero, we are sure that no query was bucket suppressed, and hence all $q_j$'s are valid samples. The test is considered passed in this case, and failed otherwise. See the algorithm \ref{proc:auto-valid-assumption-1} for an implementation example.

Validating the second assumption relies on a heuristic. We run the query:
\[
\coun(A' = x^{(A')} \wedge u = x^{(u)})
\]
and consider the assumption validated if the output is zero, and not otherwise. The idea is that if the output is larger than zero, then the query was not bucket suppressed, and many users are likely share the same attributes $x^{(A',u)}$, meaning that $x^{(A',u)}$ is unlikely to be value-unique. Experiments in section~\ref{sec:experiments} show that this heuristic works very well on real-world datasets.

\begin{procedure}
    \caption{NoBucketSuppression($A', u, x^{(A',u)}, \Delta, s, v$)}\label{proc:auto-valid-assumption-1}
	\KwIn{known attributes (names $A', u$ and values $x^{(A',u)}$), dummy conditions $\Delta$, secret $s$ and target value $v$}
	\KwOut{True if $(A',u)$ passes the tests and is deemed to satisfy assumption 1, False otherwise}
    $ok_Q \gets 0$,\ $ok_{Q'} \gets 0$  \;
    \For{$j \gets 1$ \KwTo $|\Delta|$}{
        $\varphi \gets A' = x^{(A')} \wedge \bigwedge_{l \neq j} \Delta_l$ \;
        $\widetilde{Q} \gets \coun\left(\varphi \wedge s \neq v\right)$ \;
        $\widetilde{Q}' \gets \coun\left(\varphi \wedge u \neq x^{(u)} \wedge s \neq v\right)$ \;
        \If{$\widetilde{Q} > 0$}
            {$ok_Q \gets 1$}
        \If{$\widetilde{Q}' > 0$}
            {$ok_{Q'} \gets 1$}
    }
    \Return $ok_Q = 1$ \& $ok_{Q'} = 1$
\end{procedure}

\begin{procedure}
    \caption{ValueUnique($A', u, x^{(A',u)}$)}\label{proc:auto-valid-assumption-2}
	\KwIn{known attributes (names $A', u$ and values $x^{(A',u)}$)}
	\KwOut{True if $(A',u)$ passes the tests and is deemed to satisfy assumption 2, False otherwise}
     $\widetilde{Q} \gets \coun (A'=x^{(A')} \wedge u = x^{(u)})$ \;
     \Return $\widetilde{Q} = 0$
\end{procedure}

The procedures ~\ref{proc:dummy-cloning-attack} and \ref{proc:auto-valid-assumption-1} both issue (the same) $2 |\Delta|$ queries, while \ref{proc:auto-valid-assumption-2} uses only one query. Validating the assumptions and performing the attack thus requires only $2 |\Delta| + 1$ queries, for a given set of attributes $(A',u)$. We empirically obtain accuracy above 93.3\% with $|\Delta|$ as low as 10 (see section \ref{sec:experiments} and Appendix \ref{appendix-number-of-queries}).

\point{Full cloning attack}
Combining procedures~\ref{proc:dummy-cloning-attack}, \ref{proc:auto-valid-assumption-1} and \ref{proc:auto-valid-assumption-2}, we design a fully fledged procedure \ref{proc:full-cloning-attack} that performs the entire attack under the following assumptions:
\begin{itemize}
\item[H1] The attacker knows that the victim's record $x$ is in the dataset.
\item[H2] The attacker knows a set $A^*$ of the correct attributes for the victim with $|A^*|=k^*$, i.e.\ the background knowledge is $x^{(A^*)}$.
\item[H3] The secret attribute $x^{(s)}$ is a binary attribute.
\end{itemize}

The full cloning attack includes a subset exploration step similar to the one used in the full differential attack. The algorithm selects random subsets $A'$ of $A^*$ (and an element $u$ from $A^* \setminus A'$ at random) by decreasing size until it finds a subset that passes both tests, upon which it then performs the attack using $x^{(A',u)}$ as background knowledge. If no feasible subset is found, the algorithm outputs NonAttackable.

\begin{procedure}
    \caption{FullCloningAttack($A^*, x^{(A^*)}, \Delta, s, v$)}\label{proc:full-cloning-attack}
	\KwIn{known attributes (names $A^*$ and values $x^{(A^*)}$), dummy conditions $\Delta$, secret $s$ and target value $v$}
	\KwOut{True if $x^{(s)} = v$, False if $x^{(s)} \neq v$}
    \For{$k \gets |A^*|$ \KwTo $1$}{
    	\For{iter $\gets 1$ \KwTo $100$}{
    		$A' \gets \text{RandomSubsetOfSize}(A^*, k-1)$ \;
    		$u  \gets \text{RandomElement}(A^* \setminus A')$ \;
    		\If{{\upshape \ref{proc:auto-valid-assumption-1}$(A', u, x^{(A)}, \Delta, s, v)$ \normalfont{and} \ref{proc:auto-valid-assumption-2}$(A', u, x^{(A^*)})$}}
    			{\Return $\text{CloningAttack}(A', u, x^{(A)}, \Delta, s, v)$}
    	}
    }
    \Return NonAttackable
\end{procedure}

\point{Reducing the number of queries}
While Diffix allows each analyst to send arbitrarily many queries, we study how many queries are required to perform the cloning attack in practice. In Appendix \ref{appendix-number-of-queries} we present a heuristic that reduces the median number of queries by a factor of 100. Using this heuristic, the attack targets 55.4\% of the users in the dataset, achieving 91.7\% accuracy with a maximum of $\maxNQueriesImproved$ queries per user in our experiments.
\section{Experiments}
\label{sec:experiments}

In order to assess the  effectiveness of our attacks, we implemented Diffix's mechanism for counting queries as described in the original paper~\cite{Francis_2018-dv}. The implementation outputs zero when queries are bucket-suppressed and results are rounded to the nearest nonnegative integer.
We apply our attacks to four datasets and an additional synthetic dataset on which the assumptions of the differential attack are always validated.

\subsection{Description of the datasets}

In our experiments, we use the following datasets:
\begin{enumerate}
\item ADULT: U.S. Census dataset with 30,162 records and 11 attributes, incl.\ salary class as secret attribute~\cite{AdultDataset}.
\item CREDIT: credit card application dataset with 690 records and 16 attributes, incl.\ accepted credit as secret attribute~\cite{CreditDataset}.
\item CENSUS: U.S. Census dataset with 199,523 records and 42 attributes, incl.\ total personal income (digitized, null income as negative condition) as secret attribute~\cite{CensusDataset}.
\item CDR: synthetic collection of phone metadata with 2,000,000 records generated using real-world data for human behaviour and the geography of the UK for the location of antennas. Every user is a record of 11,674,870 binary attributes (an attribute being whether a user was geographically present at a certain place and time, and placed a call or received a text message). As the vast majority of the attributes in a record are null, the distribution of values for a random attribute is heavily skewed towards zero. To obtain a balanced experiment, for~50\% of runs we select as the secret attribute a pair $(\mathit{location}, \mathit{time})$ where the user was present, and for the other~50\% we select a pair where the user was absent.
\end{enumerate}

\subsection{Differential noise-exploitation attack}

\point{Evaluation of the attack alone}
We first test the differential attack on a synthetic dataset where all users satisfy the uniqueness assumption.

In the \textsf{Complete$_k$} dataset, every user is unique according to $k$ attributes (excluding the secret attribute), whereas $k-1$ attributes always identify a \emph{larger} set of users. This ensures that ($i$) every user is vulnerable to the attack and ($ii$) bucket suppression is unlikely to be triggered by the attack queries.
To create the dataset, we fix an integer $B$ and generate every possible $k-$tuple whose values are in $\{1,\dots,B\}$. We then append to each tuple a random value of either $0$ or $1$ for the secret attribute. \textsf{Complete$_k$} contains $B^k$ records, one for each combination of $k$ attributes. For our experiments, we set $B=12$, to ensure that close to no bucket suppression occurs. For computational reasons, as the size of the dataset in memory grows as $B^k$, the maximum $k$ we can use is limited to $6$.

\begin{figure}[!b]
\centering
\includegraphics[width=0.75\columnwidth]{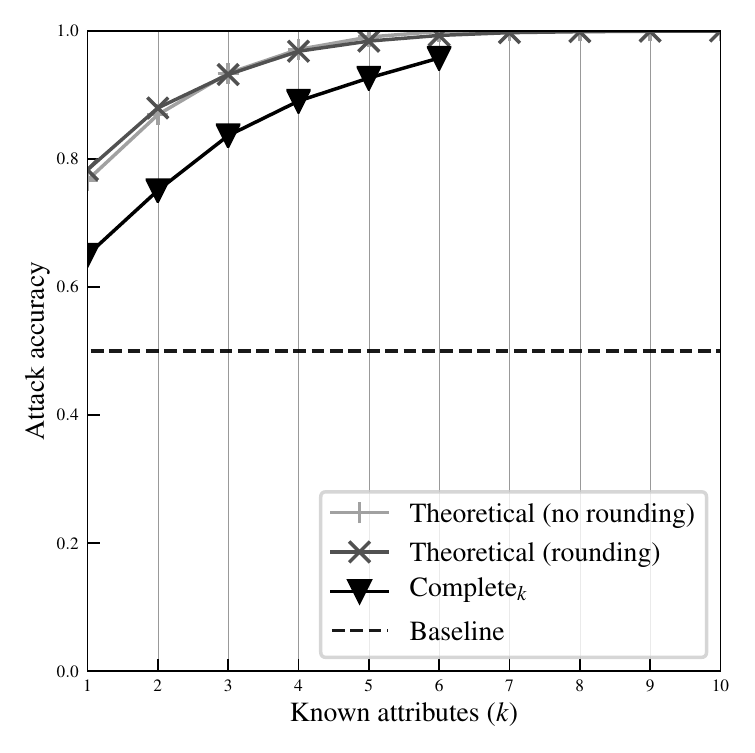}
\caption{Accuracy of the differential attack when the uniqueness assumption is always validated and with balanced truth values. The baseline accuracy is 0.5 and represents the expected success rate when randomly predicting the secret attribute using a uniform prior.}
\label{fig:likelihood-test}
\end{figure}

Fig.~\ref{fig:likelihood-test} compares the accuracy $\operatorname{acc}(k)$ of the attack, knowing $k$ attributes, on \textsf{Complete$_k$} with the theoretical accuracy. The procedure we use here is \ref{proc:differential-attack}, which does \emph{not} include subset exploration and has no access to the oracle. Hence, this experiment simulates a realistic attacker. We report the empirical fraction of users whose secret attribute is correctly predicted, estimated by performing the differential attack on a sample of 1000 users. We also report the theoretical distribution of accuracy, ($i$) without rounding (closed-form expression, see Appendix~\ref{LRT-appendix}) and ($ii$) with rounding (numerical simulation, see Appendix~\ref{LRT-appendix}).
For the \textsf{Complete$_k$} dataset, the accuracy reaches $\completekAccuracyDifferential\%$ for $\numberOfAttributesDifferential$ points. Even knowing only $k=2$ attributes, the accuracy is above 66\% both theoretically and empirically.
While rounding has close to no effect on the theoretical accuracy of the attack, comparing the \textsf{Complete$_k$} with the Theoretical (rounding) curves shows that bucket suppression and potential correlations between the samples in empirical experiments noticeably decrease the accuracy.

\point{Evaluation on real-world datasets}
We now evaluate the accuracy of the differential attack on 1,000 users selected at random in each of the four datasets.
Contrary to the synthetic experiment, bucket suppression is more prevalent on real-world datasets. Therefore, we run the \ref{proc:full-differential-attack} algorithm, knowing $k^*$ attributes $A^*$ that are selected at random for each record. If the attack outputs NonAttackable, the secret attribute is predicted at random (uniformly).

\begin{figure*}[tb]
\centering
\includegraphics[width=\textwidth]{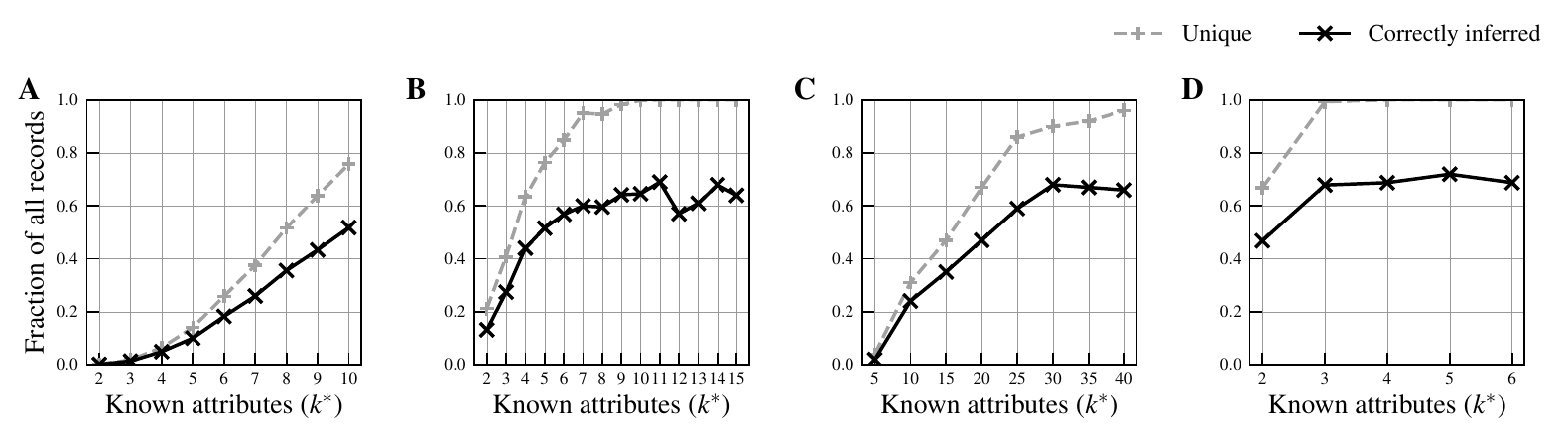}
\caption{Results of the differential attack for the (A) ADULT, (B) CREDIT, (C) CENSUS, and (D) CDR datasets.}
\label{img:results-attack1}
\end{figure*}

Fig.~\ref{img:results-attack1} shows, for each dataset and knowing $k^*$ attributes, the percentage of unique individuals and the percentage of individuals, in each dataset, for which the secret is correctly inferred. The latter divided by the former gives us the accuracy of our attack.
Our attack realizes an accuracy of 68.4\% for ADULT with $k^*=10$, 64.0\% for CREDIT with $k^*=15$, 68.8\% for CENSUS with $k^*=40$, and 68.8\% for CDR with $k^*=6$.

Observe that the fraction of correctly inferred attributes plateaus with larger $k^*$ for the CREDIT, CENSUS and CDR datasets. The reason is that, on these datasets, most users are unique for larger values of $k^*$. As explained in section \ref{attack_section}, this makes bucket suppression more prevalent and reduces the total number of samples for the likelihood ratio test, which is a limitation of the differential attack.

\subsection{Cloning noise-exploitation attack}
We implement the attack as described by algorithm \ref{proc:full-cloning-attack}. As before, for each value of $k^*$ we select 1,000 users at random, and for each user a random subset $A^*$ of their attributes of size $k^*$. $A^*$ represents the total number of attributes known to the attacker about the victim.

We set a threshold $\sigma^* = 0.7$ for the variance cutoff (see Appendix~\ref{threshold-appendix}). We generate $|\Delta| = 10$ dummy conditions for $x_1=a_1$ of the form $x_1 \neq b_j$ for $j \leq |\Delta|$, with $b_j$ being some plausible values for $x_1$ different from $a_1$. We present the results when the attacker knows enough attributes ($k^*$) to identify every user, or up to all available attributes in the dataset.

Table~\ref{tab:accuracy_attack2} shows the proportion of records that are value-unique (third column) and fraction of the value-unique records that are predicted as attackable by procedures \ref{proc:auto-valid-assumption-1} and \ref{proc:auto-valid-assumption-2} (fourth column). We then perform the cloning attack on all records that are predicted as attackable and report $\accuracypvu$, the fraction of predicted attackable records whose secret attribute was successfully inferred (fifth column). For completeness, we also report $\accuracyall$, the fraction of all records in the datasets (including the ones deemed NonAttackable) whose secret attribute was successfully inferred (last column).

Table~\ref{tab:accuracy_attack2} shows that the cloning attack---including the assumption validation step---performs really well on all datasets considered, between $\minAttacked$ and $\maxAttacked\%$ of secret attributes and $\maxAttacked\%$ on the CREDIT dataset when knowing 15 attributes.

\begin{table}[tb]
\renewcommand{\arraystretch}{1.3}
\centering 
\begin{tabular}{|l|r|r|r|r|r|r|}
\hline
\bf Dataset & \rot{\bf Attributes ($k^*$) \ } & \rot{\bf Value-unique} & \rot{\parbox{2cm}{\bf Predicted \\ attackable}} &  \rot{\bf $\accuracypvu$} & \rot{\bf $\accuracyall$} \\ \hline\hline
ADULT  &  10 &   93.0\% &        96.8\% &      93.3\% &      87.0\% \\ \hline
CREDIT &  15 &  100.0\% &       100.0\% &      97.0\% &      97.0\% \\ \hline
CENSUS &  40 &   99.7\% &        94.6\% &      97.1\% &      91.6\% \\ \hline
CDR    &   6 &  100.0\% &       100.0\% &      91.3\% &      91.3\% \\ \hline
\end{tabular}
\caption{Empirical results of the cloning attack on four real-world datasets.}
\label{tab:accuracy_attack2}
\end{table}

\begin{figure*}[tb]
\centering
\includegraphics[width=\textwidth]{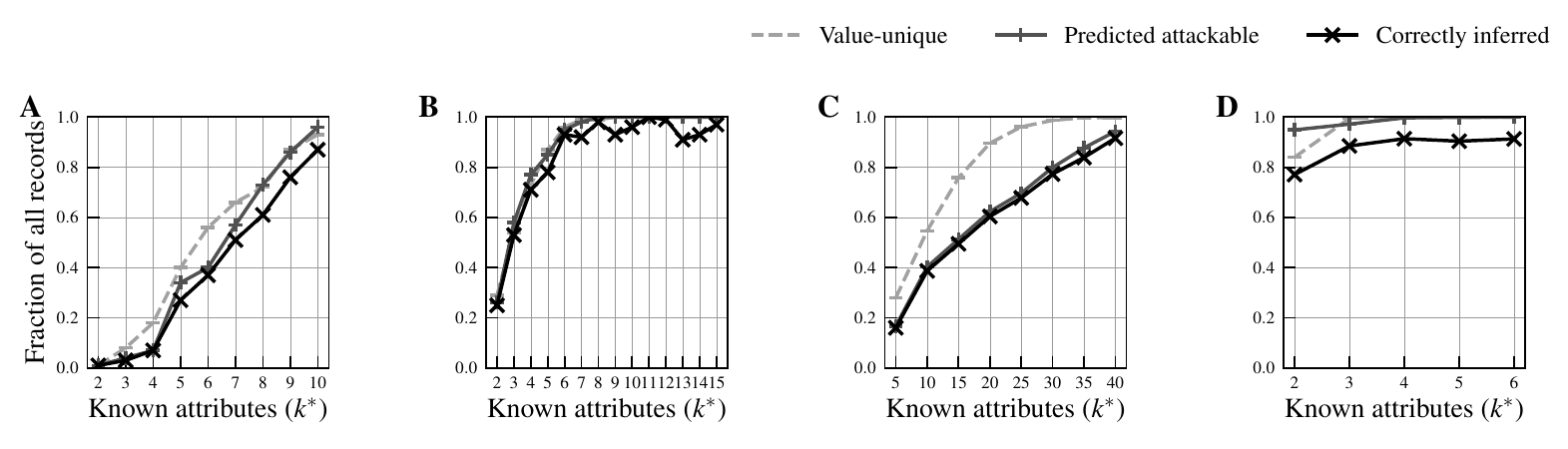}
\caption{Results of the cloning attack for the (A) ADULT, (B) CREDIT, (C) CENSUS, and (D) CDR datasets.}
\label{img:results-attack2}
\end{figure*}

Fig.~\ref{img:results-attack2} shows, knowing $k^*$ attributes, the fraction of all records that are value-unique, predicted attackable, and correctly inferred. The curves for value-unique users and for predicted attackable users are always very close, suggesting that the assumption validation step is effective. Out of all records predicted as attackable, most of them are correctly inferred, demonstrating that the attack works on targeted records across all $k$. For the {CREDIT} dataset, with only six attributes, the attack reaches the inference step for 95\% of the users in the dataset, and correctly infers the secret attribute for 93\% of the total records.

\section{Discussion}
\label{sec:discussion_section}

\subsection{Value-uniqueness and attribute predictability}\label{sec:v-uniqueness_and_predict}
Value-uniqueness plays an important role in the cloning attack. As Fig.~\ref{img:results-attack2} shows, it is a valid assumption for real-world datasets.

Value-uniqueness means that a group of people who share the same attributes also share the same secret attribute. If this group were to be large enough, the noise added by Diffix might not be enough to hide the secret attribute, which could then be revealed by using a simple count query. While this might be true for some datasets, it is not the case for any of the datasets we considered. For instance, the average size of the value-unique class (i.e.\ the set of value-unique users sharing the same restricted record) in the ADULT dataset is 1.44, with no class containing more than 4 users and similar numbers for the other datasets. This means that, most of the times, value-unique users are simply \emph{unique}.

If secret attributes are predictable from the other attributes, a trained machine learning classifier could predict them with potentially high accuracy\footnote{Whether this would constitute a privacy attack is debated~\cite{McSherry_statistical_inference}.}. Despite our datasets coming from the machine learning literature, our attack does not rely at all on the predictability of secret attributes and performs equally well if no correlation at all exists between attributes and the secret attribute. In Appendix~\ref{appendix_v-unique_and_predict}, we run our attack on a modified version of the ADULT dataset where sensitive attributes have been randomly sampled, thereby theoretically destroying any correlation. Our attacks perform as well on this modified dataset as on the original dataset.

\subsection{Producing and detecting dummy conditions}
The cloning attack requires the attacker to provide a set of dummy conditions that affect the noise addition without affecting the query set. These conditions can be syntactic (e.g., $age \geq 15$ for the query $age = 23$), semantic (e.g., $status \neq retired$ for $age = 23$), or pragmatic (e.g., $age \neq 15$ against a database containing only adult individuals).

When the language is rich enough, detecting redundant clauses is not a trivial task. At the same time, the richer the syntax is, the more utility an analyst gets out of the system. Diffix offers a fairly rich syntax including boolean expression, GROUP BY, JOIN, seven aggregation functions, set membership, fourteen string functions, and ten maths functions. In this context, automatically detecting dummy conditions would likely require iterating recursively through every condition and evaluating the query with and without them, a costly operation. Moreover, if dummy conditions are detected by evaluating them on the dataset, filtering them might not be safe. Removing semantic and pragmatic dummy conditions from a query would indeed reduce the total variance of the added noise and leak information about the dataset itself (e.g., only adult individuals are present if the condition $age \neq 15$ is always removed).

\subsection{Improving the attacks}\label{sec:improving_attacks}
The cloning attack can be modified to run a double \ref{proc:auto-valid-assumption-1} test: once with $v=0$ and once with $v=1$. If both pass and the \ref{proc:auto-valid-assumption-2} test is passed as well, the attack proceeds with the actual inference procedure \ref{proc:dummy-cloning-attack} for both $v=0$ and $v=1$. The attack then makes a guess only if both inferences return the same value, and continues with the subset exploration otherwise (deeming the user NonAttackable if no working set of attributes is found).

We found that this modified attack improves the $\accuracypvu$ figure on all datasets (e.g. from 93.3\% to 97.3\% for ADULT, all results available in Appendix~\ref{appendix-improvements}). However, double tests are more likely to fail, meaning that less users are predicted as attackable (from 96.8\% to 87\% in ADULT). Because of bucket suppression, this effect is particularly strong for datasets where the overall distribution of the secret attribute is very skewed, such as the CDR dataset where the number of predicted attackable users goes from 100\% to 8.5\%. Depending on the aims of the attacker (precision versus coverage), she might prefer the original or the double version of the cloning attack. 

\point{Other improvements}
To properly quantify the strength of our attacks, none of them use prior knowledge on the distribution of the secret attribute. In practice, an attacker might want to use this information, e.g.\ obtaining it by querying Diffix. We discuss this in Appendix \ref{appendix-improvements}. We also discuss how to generate more samples for the differential attack and outline how to generalize both attacks to infer non-binary attributes.

\subsection{Defenses}
In this section, we briefly outline some of the approaches that may be used to mitigate the effects of our noise-exploitations attacks -- and other attacks -- against Diffix and other privacy-preserving query-based systems. Overall, it is our belief that practical secure design principles apply here just as they do in many other contents. 
Specifically, privacy-preserving query-based system such as Diffix (regardless of whether they have provable guarantees or not) would benefit from a defense-in-depth approach, by monitoring the query stream for queries that are likely to lead to exploitation.

\point{Intrusion detection} 
The set of queries generated by our attacks follow a specific template. Learning this pattern may help prevent noise-exploitation attacks, as well as potentially related attacks. A more sophisticated attacker might however vary the shape of the queries and interleave them with other more natural-looking queries, including over long periods of time.

\point{Auditability} 
If the user of such a system is authenticated, then a suspicious-looking query stream can lead to temporary account suspension and further investigations of their activity, including after the fact, as new attacks are being uncovered.

\point{Increased friction}
Another strategy involves imposing time delays or financial charges on queries, for instance by charging by the number of queries, instead of using a subscription-based model. This strategy can be refined to, for instance, charge more or create longer delays for suspicious queries. This would make it more difficult to automate the inference process at scale.

\point{Limited expressiveness}
Instead of a rich syntax, the mechanism could allow only for a small set of conditions that are easier to validate. This could also include a limit to the number of conditions per query, or to the total number of conditions that may be used by the authenticated system user during a specific interval of time.
This involves a compromise between rough data summarization and fine-grained queries, and limits the utility of the system in practice.

\subsection{Disclosure}
After we discovered and prototyped our differential attack, we reached out to the authors of Diffix and shared with them our manuscript, which subsequently appeared on ArXiv.org. A week later, the authors of Diffix published a blog post on their website \cite{aircloak_response} discussing our results. While they acknowledge our attack, they claim that it is not practical as the necessary assumptions are rarely met in the datasets they analyzed.

We disagree with this claim. First, the existence of the attack, \emph{independently} of the dataset, contradicts both the spirit and the letter of GDPR's Art.\ 29 WP. Second, we showed that, albeit correct, the authors' analysis was insufficient. In this paper, we give an example of a dataset on which the differential attack is very effective, even without an oracle. Moreover, we demonstrate that there exist real-world datasets on which the necessary assumptions are met for a significant fraction of users. Third, the cloning attack, which we developed afterwards, is able to validate its assumptions automatically and performs very well on a large range of real-world datasets.

The code of our differential and cloning attacks, as well as the experiments performed in this paper, are available at \url{https://cpg.doc.ic.ac.uk/signal-in-the-noise}.

\section{Related work}\label{sec:related_work}

\point{Attacks on query-based systems}
Diffix is an example of query-based system: the individual-level (often pseudonymous) data is stored on the data curator's server. Users access the data exclusively by sending queries that only return information aggregated from several records. While this setup prevents traditional re-identification attacks~\cite{Sweeney1997-cm,Narayanan2008-mj,De_Montjoye2013-sj,De_Montjoye2015-sz,Culnane2017-vc,Ohm2010-zy}, a large range of attacks on query-based systems have been developed since the late 70's~\cite{Denning1978-qd,Beck_1980}. Most of these attacks show how to circumvent privacy safeguards (for instance, query set size restriction and noise addition) in specific setups. In 2003, Dinur et al.~\cite{Dinur_Nissim_2003} proposed the first example of an attack that works on a large class of query-based systems. In what they called a reconstruction attack, they showed that if the noise added to every query is at most $o(\sqrt{n})$, where $n$ is the size of the dataset, then an attacker can reconstruct almost the entire dataset using only polynomially many queries. Sankararaman et al.~\cite{Sankararaman_Obozinski_Jordan_Halperin_2009} realized the first formal study of tracing attacks, introducing a theoretical attack model based on hypothesis testing. While reconstruction attacks aim at inferring one or more attributes of some record in the dataset (violating the inference requirement of the Art.\ 29 WP), the goal of tracing attacks is only to determine whether the data about a certain individual (more precisely, their record) is present in the dataset. Numerous reconstruction and tracing attacks have been proposed in the literature. These attacks address different limitations of previous ones, particularly the computational time required to perform them. A recent survey from Dwork et al.~\cite{Dwork_Smith_Steinke_Ullman_2017} gives a detailed overview of attacks on query-based systems.

\point{Attacks on differential privacy} Differential privacy is a privacy guarantee that can be enforced by query-based systems. Differential privacy has been mathematically proven to be robust against a very large class of attacks~\cite{Kasiviswanathan2014-am} when used with an appropriate privacy budget $\epsilon$. However, research has shown that attacks on \emph{implementations} of differentially private systems exist. We give an overview in Appendix \ref{appendix-DP-attacks}.

\point{Differential privacy for general-purpose analytics}
Diffix was specifically created as an alternative to differential privacy to provide a better privacy/utility tradeoff for general-purpose analytics~\cite{aircloak_mydata_post, aircloak_mydata_slides}. Specifically, Diffix allows for infinitely many queries with little noise added to outputs.

General-purpose analytics usually refers to systems that allow analysts to send many queries of different type, and ideally permit to join different datasets. Some solutions based on differential privacy have been proposed, the main ones being PINQ~\cite{McSherry2009-rp}, wPINQ~\cite{Proserpio2014-la}, Airavat~\cite{Roy2010-xb}, and GUPT~\cite{Mohan2012-kx}. All of these systems however present limitations, e.g.\ simplicity of use and support for various operators that join different datasets~\cite{uber_flex}. In 2017, Johnson et al.~\cite{uber_flex} proposed a new framework for general-purpose analytics, called FLEX, developed in collaboration with Uber. FLEX enforces differential privacy for SQL queries without requiring any knowledge about differential privacy from the analyst. However, the actual utility achieved -- level of noise added -- by the current implementation of FLEX has been questioned~\cite{McSherry_uber}.

\point{Attacks on data-dependent noise}
Values of Diffix's dynamic noise for a query depend on the query set (i.e.\ the set of users selected by the query), and hence on the data. This is what allows for our noise-exploitation attack to work. Data-dependent noise, also called instance-based noise, has been shown to provide significantly better accuracy than data-independent noise~\cite{Nissim2007-pd}. However, naive implementations of data-dependent noise can leak information about the data, a result Nissim et al.\ theorized as a potential way to attack the system~\cite{Nissim2007-pd}. To the best of our knowledge, our noise-exploitation attack is the first instance of an attack exploiting specifically data-dependent noise on deployed systems.

\subsection{Other attacks on Diffix} \label{sec:other-attacks}
We published the first version of our paper on ArXiv.org in April 2018, describing the differential attack. We updated it with a cloning attack in July 2018. Two months later, in October 2018, two other attacks on Diffix were disclosed. A membership attack by Pyrgelis et al.~\cite{LocationTimeMembership}, based on a previous paper\cite{pyrgelisKnockKnockWho2018}, and a reconstruction attack by Cohen and Nissim~\cite{cohenLinearProgramReconstruction2018}, based on previous work by Dinur et al.~\cite{Dinur_Nissim_2003} and Dwork et al.~\cite{dworkPricePrivacyLimits2007}. These attacks are very different from ours and require a large number of queries in a typical setting, while our cloning attack can work with only 32 queries (see Appendix \ref{appendix-number-of-queries}). These are, to the best of our knowledge, the only three attacks specifically targeting Diffix.

\point{Membership attack on location data}
The attack by Pyrgelis et al.~\cite{LocationTimeMembership} is as follows: the attacker trains a machine learning algorithm on a \emph{linkability} dataset (the attacker's background knowledge) to infer the presence of a user in a \emph{protected} dataset (accessible only through queries on Diffix). Both datasets contain the full trajectories of users and half of the users are present in both datasets. The classifier is trained on the linkability dataset and queries on Diffix that count the number of people transiting in a certain area at a given time. The experimental results focus on the top 100 users with the highest number of reported locations in the linkability dataset. Out of 62 users present in both datasets, the classifier correctly infers the presence for 50 of them.

This attack presents three limitations. First, it is a membership attack and only allows an attacker to infer whether a person is in the protected dataset or not. Second, it assumes a strong adversary who has access to the full trajectory of a user exactly as it exists in the protected dataset, for a large number of users. Third, the attack requires about 32,000 queries to assess the presence of a user. Membership attacks are however very useful when combined with inference attacks like ours, allowing an adversary to effectively verify our assumption that the victim is in the dataset.

\point{Linear program reconstruction attack}
The attack by Cohen et al.~\cite{cohenLinearProgramReconstruction2018} focuses on reconstructing the dataset. In its simplest form, the attack assumes that the dataset contains $n$ records, and each user $i \in [n]$ has a binary attribute $s_i$. The attack then selects random subsets of users and, for each subset $I \subseteq [n]$, queries Diffix for the result of $\sum_{i \in I} s_i$. This allows the attacker to produce a noisy linear system that can be solved using linear programming techniques to reconstruct the entire set of secret attributes $\{s_1,\ldots,s_n\}$ with perfect accuracy in polynomial time.

While this attack can successfully reconstruct the entire dataset, it presents two limitations compared to our attack. 

First, it requires that the system allows queries of the type $\sum_{i \in I} s_i$, i.e.\ queries that select any analyst-defined set of users $I \subseteq [n]$, the ``row-naming problem''. The authors here exploit SQL functions supported by Diffix to define hash functions which they then use to select ``random enough'' sets of users. Following the disclosure, Aircloak restricted the available SQL functions to prevent the attack \cite{FixMITGeorgetown2018}.

Second, to target a specific user, the attack would require a number of queries proportional to the number of records. Since the attacker does not know \emph{which} name $i \in [n]$ corresponds to the victim's record, it is necessary to fully reconstruct at least a few columns entirely. The attacker would then perform a uniqueness attack on the reconstructed dataset to infer the secret attribute of the victim. 

On the contrary, the number of queries used by our attacks is independent of the number of records in the dataset.
\section{Conclusion}\label{sec:conclusion}
The Diffix mechanism has recently been proposed as an alternative to data anonymization methods and differential privacy, and is currently used in production. The mechanism is claimed to allow an analyst to submit an unbounded number of queries, while thwarting inference attacks, as defined by EU's Art.~29 WP.
In this paper, we show that Diffix's anonymization mechanism is vulnerable to a new class of attacks, which we call noise-exploitation attacks. Our attacks leverage design flaws in Diffix's data-dependent noise to infer private attributes of an individual in the dataset, solely from prior knowledge about other attributes of this individual. In our opinion, Diffix alone and in its present state likely fails to satisfy the EU's Art.~29 WP requirements for data anonymization. Furthermore, our results show that naive data-dependent noise leads to highly vulnerable systems.

Our differential noise-exploitation attack, given little auxiliary information about the victim, combines specific queries and estimates how the noise is distributed to infer the value of the private attribute. In a synthetic best-case dataset, the attacker can predict with $\completekAccuracyDifferential\%$ accuracy private attributes, using only $\numberOfAttributesDifferential$ attributes. 

Our cloning noise-exploitation attack extends the first one by adding ``dummy'' conditions that do not change the selected query set. It relies on weaker assumptions, that are automatically validated with high accuracy by our algorithm. We evaluate its performances on four real-world datasets and find that it infers private attributes of between $\minAttacked\%$ and $\maxAttacked\%$ of all records across datasets.

We finally recommend four defense-in-depth principles to defeat the de-anonymization attacks we describe.

{\normalsize \bibliographystyle{unsrt}
\bibliography{diffix_main_conference}}

\appendix

\section{Likelihood ratio test} \label{LRT-appendix}
Let $X$ and $Y$ be independent random variables. Suppose that we have two hypotheses about the distributions of $X$ and:
\begin{align*}
H_0 &: \quad X \sim \N(\mu_0,\sigma_0^2) \quad \text{and} \quad Y \sim \N(\mu_1,\sigma_1^2) \\
H_1 &: \quad X \sim \N(\mu_1,\sigma_1^2) \quad \text{and} \quad Y \sim \N(\mu_0,\sigma_0^2)
\end{align*}
where $\mu_0, \mu_1, \sigma_0, \sigma_1$ are known and fixed values such that $\mu_0 < \mu_1$ and $\sigma_0^2 < \sigma_1^2$. $\N(\mu,\sigma^2)$ denotes the the normal distribution with mean $\mu$ and variance $\sigma^2$. 

Suppose we have a vector of $n$ realizations $\vec{x} = (x_1,\ldots,x_n)$ of $X$ and a vector of $n$ realizations $\vec{y} = (y_1,\ldots,y_n)$ of $Y$. We assume that all the $2n$ realizations are mutually independent. The standard frequentist way to accept the preferred hypothesis $H_0$ or refute it (in favor of $H_1$) would use a likelihood ratio test with a pre-defined confidence level from which to derive critical regions \cite{Young_Smith_2005}. In our case we do not have a preferred hypothesis, and hence we define a slightly different test.

Let $f$ and $g$ denote the probability density functions of $\N(\mu_0,\sigma_0^2)$ and $\N(\mu_1,\sigma_1^2)$ respectively. We define the likelihood ratio function $\Lambda$ as follows:
\begin{equation*}
\Lambda(\vec{x}, \vec{y}) = \prod_{j=1}^n \frac{f(x_j)}{g(x_j)} \prod_{j=1}^n \frac{g(y_j)}{f(y_j)}.
\end{equation*}
We accept $H_0$ if $\Lambda(\vec{x}, \vec{y}) \geq 1$, and we accept $H_1$ if $\Lambda(\vec{x}, \vec{y}) < 1$.

\point{Theoretical accuracy of the test}
The test will sometimes yield the wrong result. It is possible to determine what is the probability that this happens. Such probability depends on mean and variance of the two specific normal distributions.
\begin{fact}
Let $p_{\text{err}_0} = \Pr[\Lambda(\vec{x}, \vec{y}) < 1 \mid H_0]$ and $p_{\text{err}_1} = \Pr[\Lambda(\vec{x}, \vec{y}) \geq 1 \mid H_1]$. Then
\begin{equation*}
p_{\text{err}_0} = p_{\text{err}_1} = \Pr[\alpha_0 Z_0 - \alpha_1 Z_1 < 0]
\end{equation*}
where, for $i=0,1$, $Z_i$ is a noncentral chi-squared distribution with $n$ degrees of freedom and noncentrality parameter
\begin{equation*}
\lambda_i = n \left( \frac{\mu_i}{\sigma_i} + \frac{\mu_0 \sigma_1^2 - \mu_1 \sigma_0^2}{\sigma_i (\sigma_0^2 - \sigma_1^2)} \right)^2
\end{equation*}
and
\begin{equation*}
\alpha_i = \frac{\sigma_0^2 - \sigma_1^2}{2\sigma_{1-i}^2}.
\end{equation*}
\end{fact}
To prove the fact, one considers the log-likelihood ratio function $\log \Lambda(\vec{x}, \vec{y})$ and applies elementary algebra to the obtained expression to derive a linear combination of noncentral chi-squared distributions. We omit the details.

Since $p_{\text{err}_0} = p_{\text{err}_1}$, we refer to this quantity simply as $p_{\text{err}}$. The accuracy of the test is $\operatorname{acc} = 1-p_{\text{err}}$.

We now show how to apply the fact to our differential noise-exploitation attack. For simplicity, we suppose that Diffix's outputs are not rounded to the nearest nonnegative integer and bucket suppression is never triggered for the queries in the attack, so that every pair of queries $(\widetilde{Q}_j,\widetilde{Q}'_j)$ and $(\widetilde{R}_j,\widetilde{R}'_j)$ yields a valid sample. Thus, for $k$ known attributes, we have two vectors of samples $\vec{q} = (q_1,\ldots,q_k)$ and $\vec{r} = (r_1,\ldots,r_k)$ and for every $j \leq k$:
\begin{equation*}
q_j \sim \begin{cases}
  \N(0,2) & \text{ if } x^{(s)} = 1 \\
  \N(1,2k+2) & \text{ if } x^{(s)} = 0
\end{cases}
\end{equation*}
\begin{equation*}
r_j \sim \begin{cases}
  \N(1,2k+2) & \text{ if } x^{(s)} = 1 \\
  \N(0,2) & \text{ if } x^{(s)} = 0
\end{cases}
\end{equation*}
We assume that the $2k$ samples in $\vec{q}$ and $\vec{r}$ are mutually independent. As discussed in section \ref{attack_section}, this is not always guaranteed to be true, but it has close to no effect on the actual accuracy of the test. Let
\begin{align*}
H_0 &: \quad x^{(s)} = 1 \\
H_1 &: \quad x^{(s)} = 0.
\end{align*}
Let $f$ and $g$ denote the probability density functions of $\N(0,2)$ and $\N(1,2k+2)$ respectively. Observe that $H_0$ holds if and only if every $q_j \sim f$ and every $r_j \sim g$. Similarly, $H_1$ holds if and only if every $q_j \sim g$ and every $r_j \sim f$. Then we can apply the test defined above. Define
\[
\Lambda(\vec{q}, \vec{r}) = \prod_{j=1}^k \frac{f(q_j)}{g(q_j)} \prod_{j=1}^k \frac{g(r_j)}{f(r_j)}.
\]
Our test concludes that $x^{(s)} = 1$ if $\Lambda(\vec{q},\vec{r}) \geq 1$, and $x^{(s)} = 0$ if $\Lambda(\vec{q},\vec{r}) < 1$.

To measure the theoretical accuracy of the attack for $k$ known attributes, we can apply the fact to $\Lambda(\vec{q},\vec{r})$ with $\mu_0 = 0, \sigma_0^2 = 2, \mu_1 = 1, \sigma_1^2 = 2k+2$ and $n = k$, and finally find $\operatorname{acc}(k) = 1-p_{\text{err}}(k)$.

Fig.~\ref{fig:likelihood-test} shows the values of $\operatorname{acc}(k)$ for increasing values of $k$. Computing the value of $p_{\text{err}}$ requires an approximation of the cumulative distribution function of a linear combination of noncentral chi-squared distributions, for which an exact closed-form expression is not known \cite{Sankaran_1963}. We compute these values using the R package \texttt{sadists}\footnote{\url{https://github.com/shabbychef/sadists}} version 0.2.3.

\point{Numerical simulation with rounding}
If we suppose that Diffix's outputs are rounded to the nearest nonnegative integer, no simple expression can be determined for the error rate. To estimate the accuracy in this case, we numerically simulate the values of $\widetilde{Q}_j(D)$ and $\widetilde{Q}'_j(D)$ that would result from querying Diffix (without bucket suppression), for different values of the secret attribute $x^{(s)}$. We then obtain each sample as the difference of the rounded results: 
\[
q_j = \round(\widetilde{Q}_j(D)) - \round(\widetilde{Q}'_j(D)).
\]
Finally, we perform the likelihood ratio test as for the continuous case (considering also null outputs) and check whether the result is correct. We use balanced truth values for $x^{(s)}$, and perform 1000 experiments (on different queries) for each value of $x^{(s)}$. The results are shown in Fig.~\ref{fig:likelihood-test}.

\section{Cutoff threshold for the cloning attack} \label{threshold-appendix}

The cloning attack tests whether the empirical variance of $|\Delta|$ samples is higher than a cutoff threshold $\sigma^*$. In order to choose this threshold $\sigma^*$, we simulate the noises and samples that would result from Diffix queries. We use the expressions resulting from equations (\ref{eqn_Q1_atk2_new}) and (\ref{eqn_Q2_atk2_new}), introducing the rounding operator $\round(\cdot)$ that rounds to the nearest nonnegative integer. 

For $j = 1, \dots, |\Delta|$, we have:
\begin{align*}
\widetilde{Q}_j(D) = &\round\Big(Q_j(D)  \\
&+ \sum_{i=1, i\neq j} \static[\Delta_i] + \sum_{i=1, i\neq j} \dynamic[\Delta_i] \\
&+ \static[A' = x^{(A')} \land s=0] \\
&+ \dynamic[A' = x^{(A')} \land s=0]\Big)
\end{align*}
and, with $q = Q_j(D)$ if $x^{(s)} = 1$ or $q = Q_j'(D)$ otherwise:
\begin{align*}
\widetilde{Q}_j'(D) =& \round\Big( q \\
& + \sum_{i=1, i\neq j} \static[\Delta_i] + \sum_{i=1, i\neq j} \dynamic[\Delta_i] \\
&+ \static[A' = x^{(A')} \land s=0] \\
&+ \dynamic[A' = x^{(A')} \land s=0] \\
&+ \static[u \neq x^{(u)}] + \dynamic[u \neq x^{(u)}] \Big)
\end{align*}

We perform a balanced experiment, simulating 1000 set of answers for $x^{(s)} = 1$ and 1000 set of answers for $x^{(s)} = 0$. For each sample, we generate the different static and dynamic noises once by sampling each from $\N(0,1)$, and use these values to compute the pairs $(\widetilde{Q}_j, \widetilde{Q}_j')$. Without loss of generality, since we only measure the variance, we fix $Q_j(D) = Q_j'(D) + 1$. We then compute the sample variance of $\left(\widetilde{Q}_1-\widetilde{Q}_1', \dots, \widetilde{Q}_{|\Delta|}-\widetilde{Q}_{|\Delta|}'\right)$. 

\begin{figure}[tb]
\centering
\includegraphics[width=.85\columnwidth]{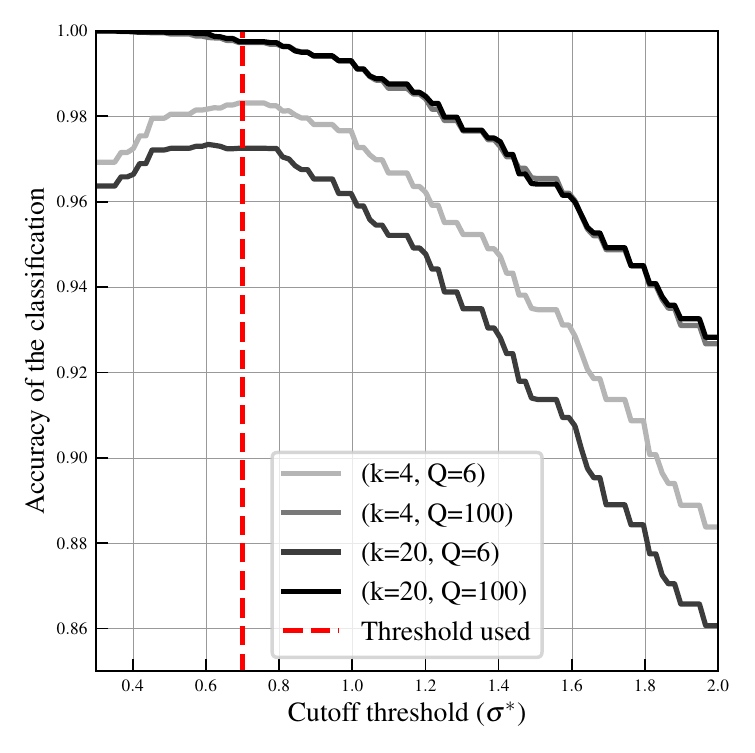}
\caption{Influence of the variance cutoff $\sigma^*$ on the accuracy of the classification for the cloning attack with $|\Delta| = 10$. The samples are obtained by simulating Diffix results $(Q_j, Q_j')$ from their distribution, using different values of the number of attributes $k$ and the true user set size $Q$.}
\label{fig:variance-cutoff}
\end{figure}

Fig.~\ref{fig:variance-cutoff} shows the accuracy at correctly predicting whether $x^{(s)} = 1$ or $x^{(s)} = 0$ from the variance of the $|\Delta|$ pairs. We find that, for $|\Delta|=10$ (10 dummy queries) and a query set of size 10, selecting $\sigma^* = 0.7$ leads to a 98.2\% true positive rate for a 98.4\% true negative rate despite rounding. This choice is arbitrary, and an attacker could choose a different value for different true and false positive rates. We however argue that the attack is robust to the choice of $\sigma^*$: while 0.7 seems to overall maximize accuracy, values of the threshold between 0.4 and 1 have comparable accuracy.

\section{Size of value-unique classes and cloning attack on randomized dataset}\label{appendix_v-unique_and_predict}

\begin{figure*}[tb]
\centering
\includegraphics[width=\textwidth]{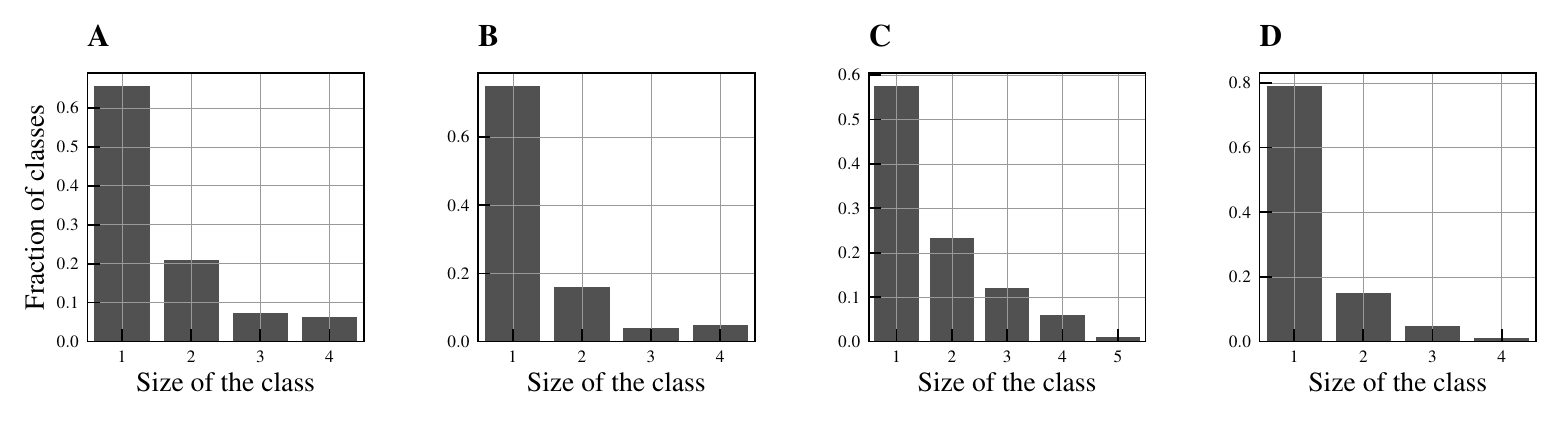}
\caption{Size of value-unique classes for users predicted attackable by the cloning attack (using all attributes) on the (A) ADULT, (B) CREDIT, (C) CENSUS, and (D) CDR datasets.}
\label{img:distribution-userset-size}
\end{figure*}

\point{Size of value-unique classes}
Fig.~\ref{img:distribution-userset-size} shows the distribution of the size of value-unique classes of users predicted attackable by the cloning attack (using all attributes). In most cases, the restricted record used by the cloning attack uniquely identifies the victim and the victim's value-unique class never contains more than 5 users.

\begin{figure}[tb]
\centering
\includegraphics[width=.85\columnwidth]{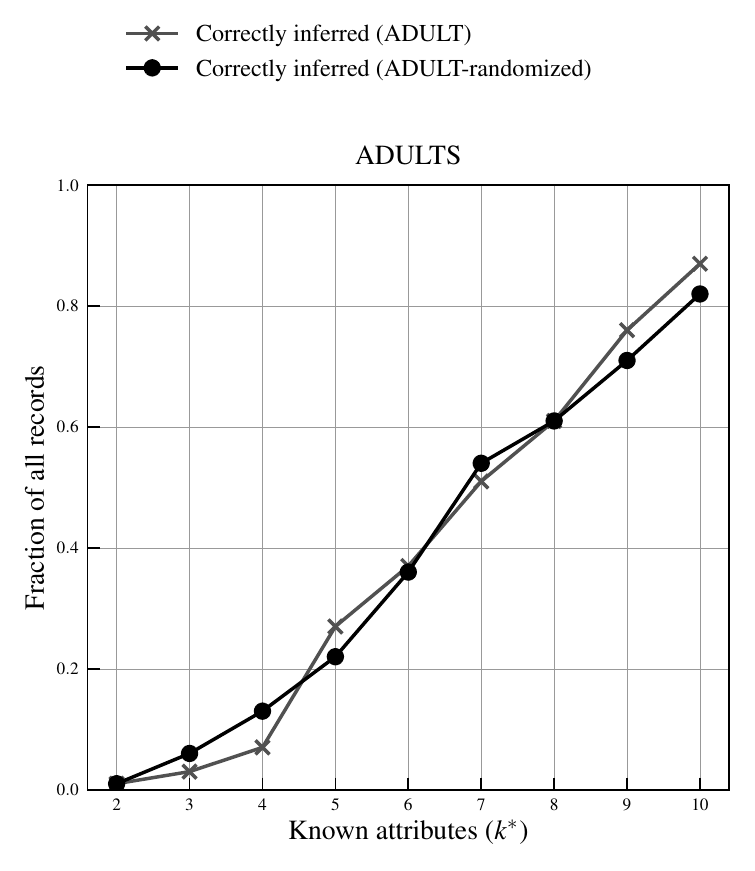}
\caption{Results of the cloning attack for the ADULT and the ADULT-randomized datasets.}
\label{img:all-attack2-ADULTS-randomized}
\end{figure}

\begin{figure}[tb]
\centering
\includegraphics[width=.85\columnwidth]{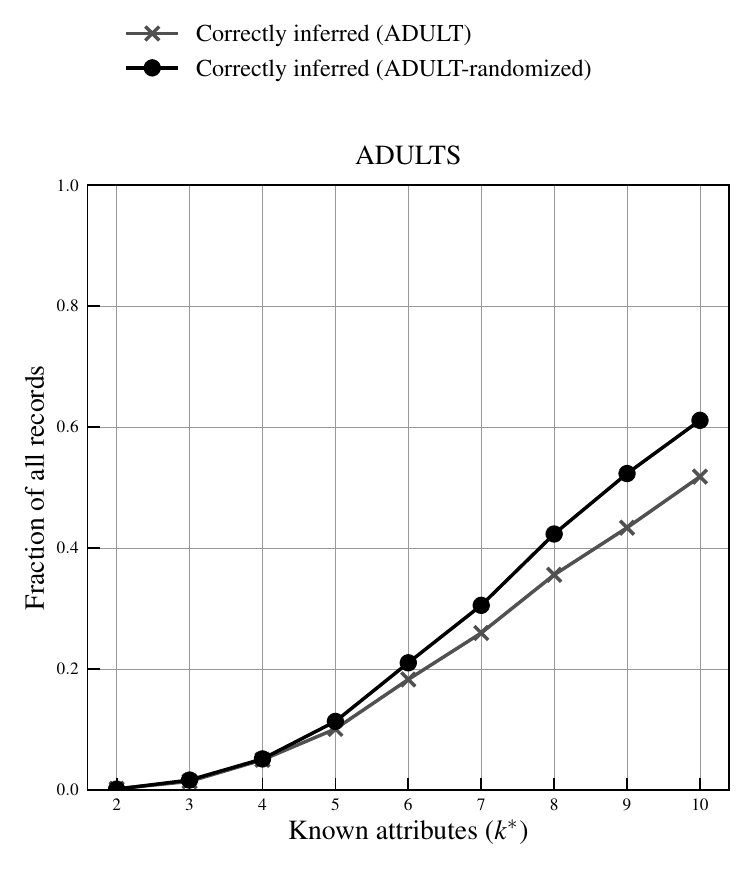}
\caption{Results of the differential attack for the ADULT and the ADULT-randomized datasets.}
\label{img:all-attack1-ADULTS-randomized}
\end{figure}

\point{Randomized dataset}
Our attacks do not rely at all on the predictability of secret attributes and perform equally well if no correlation exists between attributes and the secret attribute. We create a modified version of the ADULT dataset where the secret attributes have been randomly sampled (0 or 1 with equal probability), which we call ADULT-randomized. Fig.~\ref{img:all-attack2-ADULTS-randomized} shows that our cloning attack achieves roughly the same accuracy on both datasets, and similarly for the differential attack (Fig.~\ref{img:all-attack1-ADULTS-randomized}).

\section{Improving the accuracy and generalizing to non-binary attributes} \label{appendix-improvements}

\point{Double cloning attack}
The cloning attack can be modified using a double test, as explained in section \ref{sec:improving_attacks}. We implement this attack and perform it on all four datasets. The results are reported in Table~\ref{tab:accuracy_attack2_double}. Comparing these results with the ones for the single cloning attack (Table~\ref{tab:accuracy_attack2}), we observe that $\accuracypvu$ is slightly higher on all datasets, but less users are predicted attackable on all but the CREDIT dataset. The issue is particularly evident on the CDR dataset, where the figure drops from 100\% to 8.5\%. The reason is that the distribution of the secret attribute in the CDR dataset is extremely skewed towards 0. This means that the part of the attack using $v=1$ is likely to fail, due to many queries being bucket suppressed.

\begin{table}[tb]
\renewcommand{\arraystretch}{1.3}
\centering 
\begin{tabular}{|l|r|r|r|r|r|r|}
\hline
\bf Dataset & \rot{\bf Attributes ($k^*$) \ } & \rot{\bf Value-unique} & \rot{\parbox{2cm}{\bf Predicted \\ attackable}} &  \rot{\bf $\accuracypvu$} & \rot{\bf $\accuracyall$} \\ \hline\hline
ADULT  &  10 &   87.0\% &       86.2\% &     97.3\% &     74.0\% \\ \hline
CREDIT &  11 &  100.0\% &      100.0\% &     99.0\% &     99.0\% \\ \hline
CENSUS &  45 &  100.0\% &       87.0\% &     98.9\% &     86.0\% \\ \hline
CDR    &   6 &  100.0\% &        8.5\% &     94.1\% &      8.0\% \\ \hline
\end{tabular}
\caption{Empirical results of the double cloning attack, using all attributes.}
\label{tab:accuracy_attack2_double}
\end{table}

\point{Prior knowledge}
The differential attack and the cloning attack do not assume any prior knowledge about the distribution of the attribute $s$ in the population (i.e.\ in the dataset). In some cases, one outcome of $s$ could be much more likely than the other. For example, if the secret attribute is $\textit{HIV}$ and the dataset is randomly sampled from the population, it is more likely to observe $\textit{HIV}=0$ rather than $\textit{HIV}=1$. This prior knowledge about the population can be obtained empirically by querying the dataset itself, using $\coun(\textit{HIV}=0)$. This test can be easily integrated in the likelihood ratio test using Bayes' rule. For simplicity, we do not employ this improvement and we rather use an uninformative (uniform) prior.

\point{Obtaining more samples for the differential attack}
The differential attack relies on a likelihood ratio test with $2k$ samples, where $k$ is the number of uniquely identifying attributes. There exist several ways to generate even more samples and thus increase the attack accuracy, useful when $k$ is small or too many queries are bucket suppressed.

Consider again equations (\ref{eq:Qj}) and (\ref{eq:Q'j}) for queries $Q_j$ and $Q'_j$. The main property of $Q_j$ and $Q'_j$ is $Q_j(D) - Q'_j(D) = Q(D)$, where $Q(D)$ is the value we want to infer. There are many other pairs of queries that satisfy the same requirement and will be processed independently. For example, one could define $P_j(D)$ by adding the condition $a_j \leq x_j$ to $Q_j$ and $P'_j(D)$ by replacing $a_j \neq x_j$ with $a_j < x_j$ in $Q'_j$. The record $x$ is still the only user that can differ from one query to the other, and thus $P_j(D) - P'_j(D) = Q(D)$. By an argument similar to the one for $Q_1$ and $Q'_1$, we obtain:
\begin{align*}
\widetilde{P}_j(D) - \widetilde{P}'_j(D) \sim \begin{cases}
  \N(0,4) & \quad \text{ if } x^{(s)} = 1 \\
  \N(1,2k+2) & \quad \text{ if } x^{(s)} = 0.
\end{cases}
\end{align*}
Repeating this for $j=1, \dots, k$ yields $2k$ additional samples. Note that, as we add one condition, this creates two independent noises which will not cancel out. We can furthermore repeat the same procedure, inverting the inequalities to obtain $2k$ additional samples, leading to a total of $6k$ samples.

In general, $2k$ samples can be obtained from any pair of queries that define a partitioning attack for $Q$; different mathematical operators could also be employed to construct such partitions. Such samples will be independent most of the time. One could further exploit Diffix's rich SQL syntax by writing conditions with different syntax but identical meaning. For example, we could replace every condition ``$a_i = x_i$'' with ``\texttt{$a_i$ IN ($x_i$)}'', and similarly ``$a_i \neq x_i$'' with ``\texttt{$a_i$ NOT IN ($x_i$)}''. Since both static and dynamic noise layers depend on the string that defines the condition, changing the SQL expression produces independent noise values.

\point{Extending the attacks to non-binary attributes}
The differential and cloning attacks, as presented, assume that the secret attribute is binary. This is a common assumption in the literature \cite{Dinur_Nissim_2003, Dwork_Smith_Steinke_Ullman_2017}, as it is generally possible to extend the attacks to non-binary attributes. We now outline, using a bisection search, how to adapt our noise-exploitation attacks for secret attributes that take values in the set of real numbers (or any set with a defined order). First, we modify our noise-exploitation attacks to test whether the secret attribute is smaller or larger than a certain starting value $t_0$. To do this, it is sufficient to change the condition $s=0$ to $s \leq t_0$ (or $s \geq t_0$) in each attack query (\ref{eq:Qj}), (\ref{eq:Q'j}), (\ref{eqn_Q1_atk2_new}), (\ref{eqn_Q2_atk2_new}). We then find the correct value (up to the desired accuracy) using a bisection search that iterates the attack with the right condition $s \leq t_i$ or $s \geq t_i$. Since the attack does not achieve perfect accuracy, each step is associated to a probability of success. This stochastic root-finding search can be solved with a probabilistic bisection algorithm (see e.g.\ Frazier et al.~\cite{frazier2016probabilistic}).

\section{Reducing the number of queries} \label{appendix-number-of-queries}

One of the main features of Diffix is that it allows analysts to send an unlimited amount of queries. Many privacy attacks work by issuing a relatively large number of queries (see also \ref{sec:other-attacks}). Limiting the number of queries allowed by Diffix would thwart or significantly affect these attacks. While our actual attack procedures require a small number of queries ($2 |\Delta| + 1$ for the cloning attack), the subset exploration step can sometimes explore many sets of attributes before finding an exploitable one. To minimize the number of queries, we replace the iterative exploration with a greedy heuristic that selects only one subset which is likely to work. We focus only on the cloning attack, as it does not require an oracle and achieves much better accuracy.

The cloning attack requires a set of attributes $(A',u)$, where the restricted record $x^{(A',u)}$ uniquely identifies the victim, but the vector $x^{(A')}$ is shared across a larger population (to avoid bucket suppression). The \ref{proc:full-cloning-attack} starts with a larger set of attributes $A^*$ and iteratively explores subsets of $A^*$ to find a candidate $(A',u)$. We replace this iterative process with a single deterministic step.

Intuitively, we want $u$ to be as discriminative as possible, while for the attributes in $A'$ to select as many users as possible. This is what the procedure \ref{proc:select-subset} does. First, it computes the (approximate) fraction of users that share the same value $x^{(a)}$, for each attribute $a \in A^*$. Then it selects as $u$ the attribute associated with the lowest fraction. Now suppose that $N$ is the estimated total number of users in the dataset. The set $A'$ is selected as the smallest set of attributes associated with the highest fraction, additionally requiring that the product of all the fractions for $(A',u)$ is smaller than $1/N$. This ensures that, with high probability, the victim is uniquely identified by $x^{(A',u)}$.

\begin{procedure}
    \caption{GreedySelectSubset($A^*, x^{(A^*)}, s, v$)}\label{proc:select-subset}
	\KwIn{known attributes (names $A^*$ and values $x^{(A^*)}$), secret $s$ and target value $v$}
	\KwOut{a set of attributes $(A',u) \subseteq A^*$}
    $N \gets \coun()$ \tcp{approx.\ tot.\ number of users}
    \ForEach{$a \in A^*$}{
				$C_a \gets \coun(a = x^{(a)} \wedge s \neq v)$ \;
				$\rho_a \gets \frac{C_a}{N}$
    }
    $\{\rho_1,\ldots,\rho_{|A^*|}\} \gets$ $\operatorname{SortDescendingOrder}(\{\rho_a\}_{a \in A^*})$ \;
    $u \gets a_{|A^*|}$ \; 
    $i \gets 1$, \quad $A' \gets \emptyset$ \;
    \While{$\rho_u \prod_{a_i \in A'} \rho_{a_i} > \frac{1}{N}$}{
    		$A' \gets A' \cup \{a_i\}$ \;
    		$i \gets i+1$
    }
    \Return $(A',u)$
\end{procedure}

We can modify the \ref{proc:full-differential-attack} replacing the subset exploration with this heuristic. The modified full attack is described in the \ref{proc:simple-full-cloning-attack} procedure. 

\begin{procedure}
    \caption{GreedyFullCloningAttack($A^*, x^{(A^*)}, \Delta, s, v$)}\label{proc:simple-full-cloning-attack}
	\KwIn{known attributes (names $A^*$ and values $x^{(A^*)}$), dummy conditions $\Delta$, secret $s$ and target value $v$}
	\KwOut{True if $x^{(s)} = v$, False if $x^{(s)} \neq v$}
		$(A',u) \gets$ \ref{proc:select-subset}($A^*, x^{(A^*)}, s, v$) \;
    	\If{{\upshape \ref{proc:auto-valid-assumption-1}$(A', u, x^{(A)}, \Delta, s, v)$ \normalfont{and} \ref{proc:auto-valid-assumption-2}$(A', u, x^{(A^*)})$}}
    		{\Return $\text{CloningAttack}(A', u, x^{(A)}, \Delta, s, v)$}
    \Return NonAttackable
\end{procedure}

Observe that the \ref{proc:select-subset} procedure issues exactly $|A^*|+1$ queries. The differential attack with the assumption validation step issues at most $2 |\Delta| + 1$ queries. So, the \ref{proc:simple-full-cloning-attack} algorithm requires at most $|A^*| + 2|\Delta| + 2$ queries.

We compared the performances of the \ref{proc:full-cloning-attack} and \ref{proc:simple-full-cloning-attack} on the ADULT dataset, with the salary class as secret attribute and the other 10 attributes as $A^*$. As in section \ref{sec:experiments}, we used $|\Delta| = 10$ dummy conditions and ran the attack on 1000 random users. The results are summarized in Table \ref{tab:accuracy_attack_simple}.

\begin{table}[tb]
\renewcommand{\arraystretch}{1.3}
\centering 
\begin{tabular}{|l|r|r|r|r|r|}
\hline
\parbox{1cm}{\bf Subset \\ selection \\[1pt]} & \rot{\parbox{2cm}{\bf Median \\ n.\ queries}} & \rot{\parbox{2cm}{\bf Max \\ n.\ queries}} & \rot{\parbox{2cm}{\bf Predicted \\ attackable}} &  \rot{\bf $\accuracypvu$} \\ \hline\hline
Iterative		&  \hspace{3pt} 304 \hspace{3pt}  &	\hspace{2pt} 5310 \hspace{2pt} &        96.8\% &      93.3\%  \\ \hline
Heuristic		&  \hspace{4pt} 32 \hspace{4pt} &	 \hspace{4pt} 32 \hspace{4pt}  &       55.4\% &      91.7\%  \\ \hline
\end{tabular}
\caption{Empirical results of the cloning attack with iterative subset exploration and with the heuristic subset selection.}
\label{tab:accuracy_attack_simple}
\end{table}

The maximum (and median) number of queries used by \ref{proc:simple-full-cloning-attack} for a single user is $10 + 2 \times 10 + 2 = 32$. The median number of queries used by \ref{proc:simple-full-cloning-attack} is about 10 times higher, and the maximum is 100 times higher.

\ref{proc:simple-full-cloning-attack} effectively attacks more than half of the users, as opposed to 96.8\% of the users for the \ref{proc:full-cloning-attack}. This is due to the fact that the first attack tries a single subset of attributes per user. However, this figure is still remarkably high, given the huge reduction of required queries. Finally, the accuracy of the inference for the attacked users is almost the same.

We believe that these results give additional evidence of the power, extendability and practicability of our noise-exploitation attacks. Introducing additional optimizations, the accuracy could be improved and the number of queries could be further reduced (see Appendix \ref{appendix-improvements}).

\section{Attacks on implementations of differential privacy} \label{appendix-DP-attacks}

While differential privacy provably protects against a very large class of privacy attacks \cite{Kasiviswanathan2014-am}, the actual implementation of differentially private mechanisms can open new vulnerabilities. Haeberlen et al.~\cite{Haeberlen2011-xr} exploited covert channels~\cite{Lampson_1973} in two popular implementations of differential privacy, PINQ~\cite{McSherry2009-rp} and Airavat~\cite{Roy2010-xb}, to design state- and timing-based attacks, showing that implementation choices and bugs can introduce vulnerability, effectively disrupting the theoretical guarantees that differentially private systems are supposed to offer. Mironov showed that finite precision and rounding effects of floating-point operations can undermine the actual guarantees of differentially private mechanisms~\cite{Mironov2012-vh}. Finally, Kifer et al.\ \cite{Kifer2011-am} argued that even perfect implementations of differential privacy are vulnerable to some attacks if the attacker has very powerful background knowledge, although this is debated~\cite{McSherry_nfl}.

Attacks on implementations of differential privacy suggest that no practical system is perfectly secure against any attack. For this reason, we believe that differential privacy-based systems could benefit from defence-in-depth measures such as the ones discussed in section \ref{sec:discussion_section}.

In general, any query-based system---whether based on differential privacy or not---stores sensitive data on the curator's servers. This constitutes a risk, as an attacker could try to break into the server by exploiting vulnerabilities unrelated to the query interface. Once an attacker obtains access to the raw, potentially pseudonymized, individual-data, it is likely very easy for them to re-identify users.

\end{document}